\pdfoutput=1
\documentclass[fleqn,usenatbib,usedcolumn]{mnras}
\usepackage[british]{babel}             
\usepackage{txfonts}                    
\let\la=\lesssim 
\usepackage[T1]{fontenc}                
\usepackage{graphicx}                   
\hypersetup{pdfauthor={A. A. Orsi},
               pdftitle={The environments of high redshift radio galaxies and quasars: probes of protoclusters},
               pdfkeywords={galaxies:high-redshift -- galaxies:evolution -- methods:numerical},
               bookmarksnumbered=true}
\volume{{\rm to be submitted}}

\newcommand{\mpc}{\mbox{$ h^{-1} \rmn{Mpc}$}}
\newcommand{\lya}{\mbox{$\rmn{Ly}\alpha$}}

\newcommand{\ha}{\mbox{$\rmn{H}\alpha$}}
\newcommand{\galform}{\texttt{GALFORM}}

\newcommand{\fesc}{\mbox{$\rm{f}_{esc}$}}

\newcommand{\llya}{\mbox{$L_{\rmn{Ly}\alpha}$}}						
\newcommand{\lunits}{\mbox{$[\rmn{erg} \ \rmn{s}^{-1} \ h^{-2}]$}}		

\newcommand{\logllya}{\mbox{${\rm log}(L_{Ly\alpha}\lunits)$}}

\title[The environments of radio galaxies and quasars]
{The environments of high redshift radio galaxies and quasars: probes of protoclusters}
\author [A. Orsi et al.]
{\'Alvaro A. Orsi$^{1,2}$\thanks{Email: aaorsi@cefca.es}, Nikos Fanidakis$^3$, Cedric G. Lacey $^4$ and Carlton M. Baugh $^4$.  \\
1. Centro de Estudios de F\'isica del Cosmos de Arag\'on, Plaza de San Juan 1, Teruel, 44001, Spain.\\
2. Instituto de Astrof\'isica, Pontificia Universidad Cat\'olica, Av. Vicu\~na Mackenna 4860, Santiago, Chile. \\
3. Max Planck Institute for Astronomy, K\"{o}nigstuhl 17 D-69117 Heidelberg, Germany.\\
4. Institute for Computational Cosmology, Department of Physics, University of Durham, Science Laboratories, South Road, Durham DH1 3LE, UK.
}
\begin{document}
\maketitle
\begin{abstract}
We use the \galform\ semi-analytical model to study high density regions traced by radio galaxies and quasars at high 
redshifts. We explore the impact that baryonic physics has upon the properties of galaxies in these environments. 
Star-forming emission-line galaxies (\lya\ and \ha\ emitters) are used to probe the environments at high redshifts. 
Radio galaxies are predicted to be hosted by more massive haloes than quasars, and this is imprinted on the amplitude
of galaxy overdensities and cross-correlation functions. { We find that \lya\ radiative transfer and AGN 
feedback indirectly affect the clustering on small scales and also the stellar masses, star-formation rates and gas metallicities of 
galaxies in dense environments.}
We also investigate the relation between protoclusters associated with radio galaxies and quasars, and 
their present-day cluster descendants. The progenitors of massive clusters associated with radio galaxies and quasars 
allow us to determine an average protocluster size in a simple way. 
Overdensities within the protoclusters are found to correlate with the halo descendant masses. We present scaling relations 
that can be applied to observational data. By computing projection effects due to the wavelength resolution
of modern spectrographs and narrow-band filters we show that the former have enough 
spectral resolution to map the structure of protoclusters, whereas the latter can be used to measure the clustering 
around radio galaxies and quasars over larger scales to determine the mass of dark matter haloes hosting them.


\end{abstract}


\section{Introduction}

A fundamental ingredient of our understanding of galaxy formation and evolution is the cosmological growth of structure in the
Universe. The hierarchical growth arises from the non-linear evolution of the dark matter (DM) density field under the action of 
gravity \citep[for a review, see][]{springel06}. 
In this scenario, galaxies form in gravitational potential 
wells (i.e. dark matter haloes) that allow gas to cool and form stars, populating the cosmic web \citep{white78}. As a result, 
galaxies form and evolve in a diverse range of environments, from voids (or regions with densities well below the average), 
to field galaxies residing in average environments and to highly overdense regions. 

Observational studies of galaxy properties in different environments have shown unequivocally that their properties are 
somehow connected with the environment in which they reside 
\citep[e.g.][]{oemler74,dressler80,hashimoto98,kauffmann04}. Galaxy clusters, the most massive virialised structures in the Universe, 
have galaxy number densities of up to a few hundred times higher than the average. In such high density environments galaxies 
follow a well known morphology-density relation \citep[e.g.][]{dressler80,balogh97,goto03}. This indicates that red, early-type
galaxies are more abundant than blue, late-type galaxies in these environments. 

Different baryonic processes are thought to be responsible for the transformation of galaxies in these environments, 
such as ram-pressure stripping and tidal interactions \citep{moore96,kawata08,vandenbosch08,tecce10}. 
More recently, the 
quenching of star-formation due to AGN feedback has been proposed as a key mechanism regulating the star-formation of galaxies 
hosted by massive haloes \citep[e.g.][]{croton06, bower06,lagos08}.

From a cosmological perspective, these large structures we see in the local Universe are expected to grow from the aggregation 
of smaller haloes formed at earlier epochs \citep{press74,peebles82,davis85,lacey93}. 
The hierarchical structure formation scenario implies that massive dark matter haloes are likely to be embedded in overdense 
regions. This suggests that any galaxy property related to halo mass will also be correlated with environment on scales beyond its 
host halo, i.e. in super-halo scales \citep[e.g.][]{mo04}. This is particularly important from the perspective of the formation of 
galaxy clusters, i.e. in the
protoclusters regime \citep{overzier06}. 
These massive structures at high redshifts will eventually form a bound cluster at late
times, and thus galaxies are expected to have properties that are connected to their host halo and its surroundings.

The search for protoclusters at high redshifts can thus be a key to unveiling the role of hierarchical merging and baryonic
processes in galaxy formation and evolution. Observationally, detecting these structures has proven challenging. Common 
techniques used to detect nearby clusters, such as looking for the thermal X-ray emission from the intra-cluster medium (ICM) 
are not sensitive enough at $z\gtrsim 2$, where clusters are expected to be in the early stages of formation. Searches for 
galaxy overdensities in wide-field surveys are more effective, but unless the galaxy redshifts are confirmed spectroscopically, 
the observed overdensities are likely to be affected by redshift uncertainties that are larger than the physical size of the 
protocluster itself \citep{chiang13}. In addition, projection effects can smear out the real overdensity signal, or even enhance 
it depending on the viewing angle \citep{shattow13}. This makes the identification of overdense structures at high redshifts 
controversial.

Instead of looking for overdensities in a blind survey, it is also possible to map the environments around objects that are good 
candidates to lie in a density peak. Generally speaking, the most luminous galaxies are found to be highly clustered, meaning 
that these are hosted by massive haloes \citep[e.g.][]{norberg01}. 
High redshift quasars are among the brightest objects in the 
Universe, so it is commonly assumed that these trace the most massive structures as well 
\citep{steidel05,kashikawa07,overzier09,utsumi10,banados13,husband13}. 
Other luminous objects used to find overdensities are the so-called \lya-blobs, which can typically reach luminosities above 
$10^{43} \lunits$ over extended regions of hundreds of kpc \citep{erb11,matsuda11,uchimoto12}. A third candidate for tracers of 
massive structures are radio galaxies, which can also have typical stellar masses of the order of $10^{12} {[\rm M_{\odot}/h]}$ 
\citep{miley08,kauffmann08,donoso10,falder10, ramos-almeida13,karouzos14,hatch14}. \citet{fanidakis13} showed that, in the context of a hierarchical galaxy formation model, 
high redshift quasars are not hosted by the most massive haloes, nor are they the progenitors of 
the most massive clusters that we observe today. Instead, the model predicts that luminous radio galaxies reside in very 
massive haloes, and are thus better tracers of the most massive protoclusters. In this paper we make use of the same galaxy formation model described in \citet{fanidakis13} 
to characterise the environments around these two types of active galaxies.

In order to map the environment around a protocluster candidate the redshifts of the objects
must be known with sufficient accuracy. Typically, this is achieved with spectroscopic follow-up of a sample 
of photometric candidates, or by utilising a narrow-band filter chosen to look for a specific emission line at the redshift of 
the protocluster. Using \lya\ emitters to map the environment in a protocluster, \citet{venemans07} found that radio galaxies 
seem to pinpoint structures with masses $\sim 10^{14-15} M_{\odot} h^{-1}$ over the redshift range $2<z<5$. Likewise, 
\citet{saito14} determine an overdensity of \lya\ emitters around a radio galaxy at $z\sim 4$ that is only marginally reproduced 
in the galaxy formation model of \citet{orsi08}.

Here, we explore the properties and evolution of overdense regions by focusing specifically on those traced by radio galaxies 
and quasars in a theoretical framework.
Radio galaxies 
are expected to trace haloes that are subject to star-formation quenching due to AGN feedback in massive haloes. 
Quasars, on the other hand, are characterised by a rapid accretion of cold gas triggered by mergers or disk instabilities, which places them in a broader range of halo masses. The most luminous
quasars also experience star-formation quenching due to AGN feedback \citep{fanidakis13}.

In this paper we tackle two main problems related to these two types of active galaxies and their environments. 
First, we characterise the impact of baryonic processes on samples of emission-line galaxies populating 
overdense regions at high redshifts traced by radio galaxies and quasars.  Second, by identifying the haloes that host radio galaxies and quasars we derive a simple way to define the size of protoclusters, 
and link their overdensities to the mass of the descendant haloes at $z=0$. 


The backbone of this study is the \galform\ semi-analytical model \citep{cole00,baugh05,bower06,lagos11}. This model incorporates 
state-of-the-art prescriptions for the evolution of galaxies and their central supermassive black holes 
\citep[BH, ][]{fanidakis11,fanidakis13}, and at the same time uses a Monte Carlo radiative transfer code for \lya\ photons to 
derive physically robust \lya\ luminosities for high 
redshift galaxies \citep{orsi12}. \galform\ predicts observational properties 
of galaxies, attempting to include all relevant physical mechanisms in the galaxy formation and evolution process. Hence, 
our choice of specific tracers of high density regions at high redshifts, and the galaxy populations used to measure the 
environment around them, are all predictions that our model provides
within a robust framework, making it suitable to be confronted against observational measurements.

The structure of this paper is as follows. Section \ref{sec.models} describes the galaxy formation model used; Section 
\ref{sec.results} describes our results analysing both overdense regions and defining protoclusters traced by radio galaxies 
and quasars. Finally, Section \ref{sec.conclusions} summarises our main findings and discusses their implications. 

\section{The theoretical galaxy formation model}
\label{sec.models}

This section describes the semi-analytical model of galaxy formation used, its main features, and the modelling of active 
galaxies and emission-lines in star-forming galaxies. 

We make use of the \galform\ semi-analytical model of galaxy formation to predict the properties of galaxies as a function 
of redshift. This model is described in detail elsewhere \citep{cole00, benson03, baugh05, bower06, lagos11}. 
The variant of \galform\ used here is presented in \citet{lacey15}. 

In short, \galform\ computes the formation and evolution of galaxies in the context of the hierarchical growth of DM 
structures. The properties and merging histories of DM haloes are extracted from the {\it Millennium-WMAP7} dark-matter only 
N-body simulation.
The halo mass resolution is $\sim 10^{10}{[M_{\odot} h^{-1}]}$ and its box-side length is $500 {\rm [Mpc}\ h^{-1}]$. 
Hence, this simulation is similar to the well known {\it Millennium} run 
\citep{springel05}, except that it was run with a set of updated cosmological parameters taken from \citet{komatsu11} obtained 
using the WMAP-7 dataset,
i.e. $\Omega_b = 0.0455, \Omega_M = 0.272, \Omega_\Lambda = 0.728, n_s=0.967, \sigma_8 = 0.810$ and $h = 0.704$.

The main baryonic processes that enter in the \galform\ calculation are i) the shock-heating and radiative cooling of gas 
inside haloes leading to the formation of a disk, ii) 
quiescent star-formation in the disk, and starbursts in a galactic bulge following galaxy mergers and disk instabilities, 
iii) feedback due to supernovae, AGN and photoionisation which regulate the star-formation process, and iv) the chemical 
enrichment of the gas and stellar component. 
Galaxy luminosities are computed using a population synthesis model \citep{cole00,gonzalez-perez14}. 
Dust extinction for the stellar continuum is calculated self-consistently based on the radiative-transfer model described in \citet{lacey11}.

The variant of \galform\ used here incorporates features from different versions of the model that were used to study specific 
problems into a single model 
to provide a powerful galaxy formation tool. Most notably, the model invokes a different initial mass function (IMF) for quiescent and starburst events. A
 top-heavy IMF is used in the latter case to explain the abundance of high redshift galaxies detected in the sub-mm \citep{baugh05}. The model also
includes a treatment of star-formation in disks following the atomic and molecular hydrogen
content of the gas \citep{lagos11}, and stellar luminosities using the \citet{maraston05} 
stellar population synthesis model that incorporates the contribution 
from thermally-pulsating asymptotic giant branch (TP-AGB) stars \citep[see also][]{gonzalez-perez14}. 

\subsection{Modelling emission-line galaxies}

In order to obtain the line fluxes of several hydrogen recombination lines, \galform\ computes the total production rate 
of hydrogen ionizing photons 
(Lyman continuum photons) by integrating the composite spectral energy distribution (SED) of each galaxy over the 
extreme-UV continuum down to the Lyman break at $\lambda=912$\AA{}. Then, by assuming that 
all of these ionising photons are absorbed within the interstellar medium (ISM) of the galaxy (i.e. the escape fraction of ionising photons is set to zero), 
case B recombination is used to convert a fraction of the Lyman continuum photons into different line fluxes \citep{osterbrock89, dijkstra14}. 
This intrinsic luminosity is later adjusted for the effect of dust attenuation by computing the continuum extinction at the wavelength of the line. 

We apply the procedure outlined above to obtain the \ha\ luminosities of galaxies. In the case of \lya\, { the intrinsic luminosity of \lya\ photons 
is expected to be reduced by the scattering of \lya\ photons by neutral hydrogen atoms in the ISM and their absorption by dust grains. 
The high scattering cross-section of photons at the \lya\ line centre makes these photons undergo numerous scattering events 
with hydrogen atoms, resulting in large path lengths, and thus they are likely to be absorbed
by dust grains present in the ISM. This results in a complex radiative transfer problem 
that cannot be accurately accounted for using analytical expressions, 
except for a few idealised configurations 
\citep{harrington73,neufeld90,dijkstra06}. }

Instead, we compute the escape of \lya\ photons using a Monte Carlo radiative transfer code for \lya\ photons that allows us to 
obtain a value for the \lya\ escape fraction, \fesc, which leads to the observed \lya\ luminosity. 
A full description of the radiative transfer code and its implementation in \galform\ can be found in \citet{orsi12}. In short, 
the code follows the 
scattering, absorption and escape histories of a large number of photons (in this case, of the order of $10^3$ to $10^4$ for each galaxy). Each 
photon can 
change its direction and frequency after 
interacting with hydrogen atoms and dust grains. If the former interaction occurs, then the photon is scattered, changing its 
direction, which leads 
to a change of frequency. If the latter takes place, then the photon can be scattered or absorbed, depending on the dust albedo. 
If a photon is absorbed by a dust grain, 
then it is discarded. If a photon escapes from the medium, its final frequency is recorded. The process is repeated until an 
accurate value for the escape fraction of \lya\ photons is attained. { Following \citet{orsi12}, we restrict
the total number of photons for a given run to be at least $10^3$, and up to $5\times 10^4$ when no photon was absorbed. Hence, 
the minimum escape fraction our model can compute is $f_{\rm esc} = 2\times 10^{-5}$.}

\begin{figure*}
\raggedleft
\rotatebox{90}{\textcolor{green}{\Large \hspace{1.25cm} Radio galaxies}}
\includegraphics[width=5.7cm]{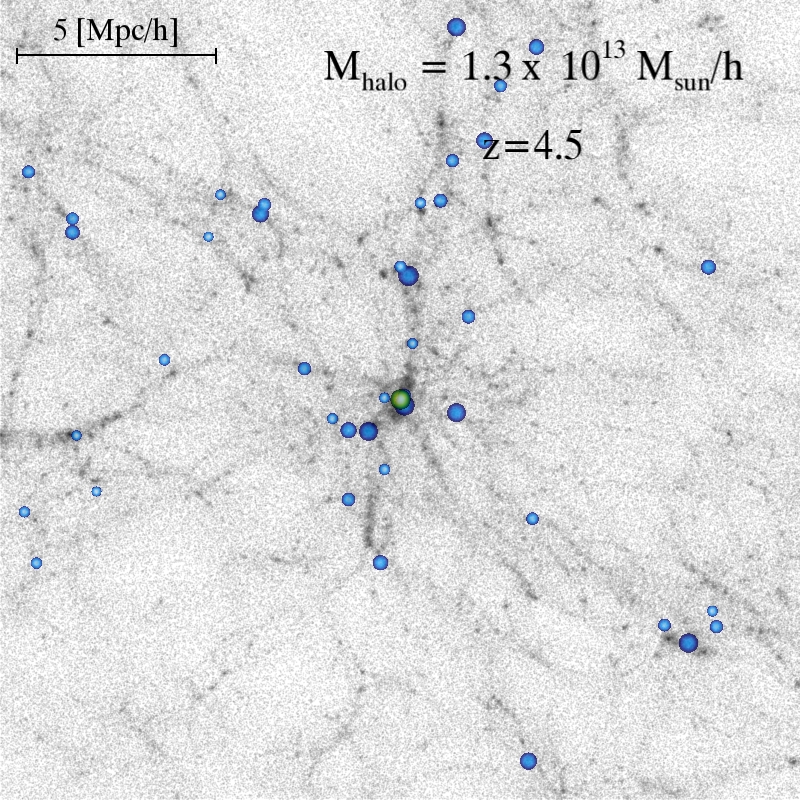}
\includegraphics[width=5.7cm]{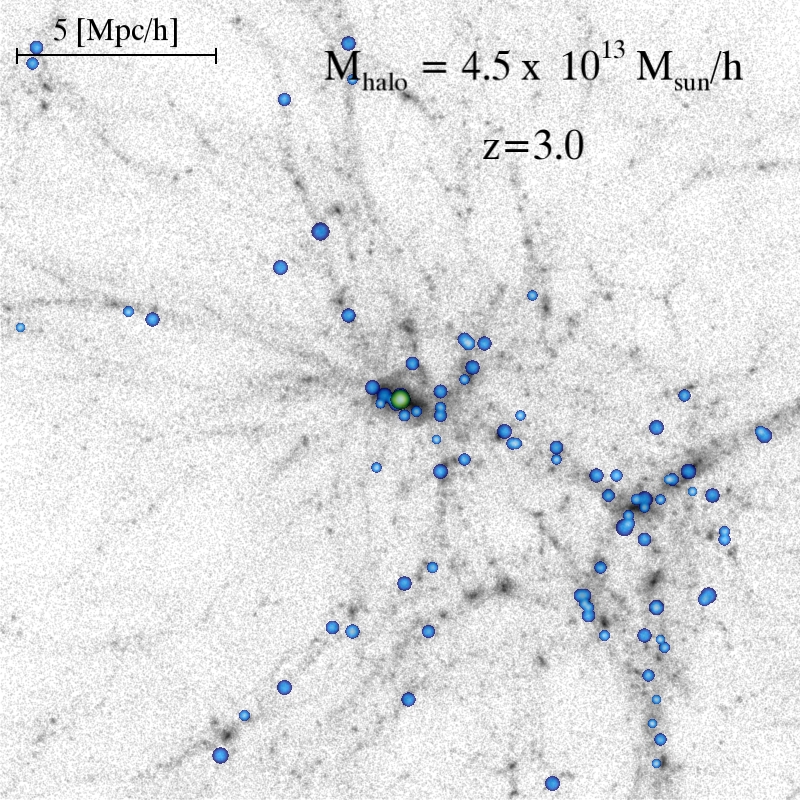}
\includegraphics[width=5.7cm]{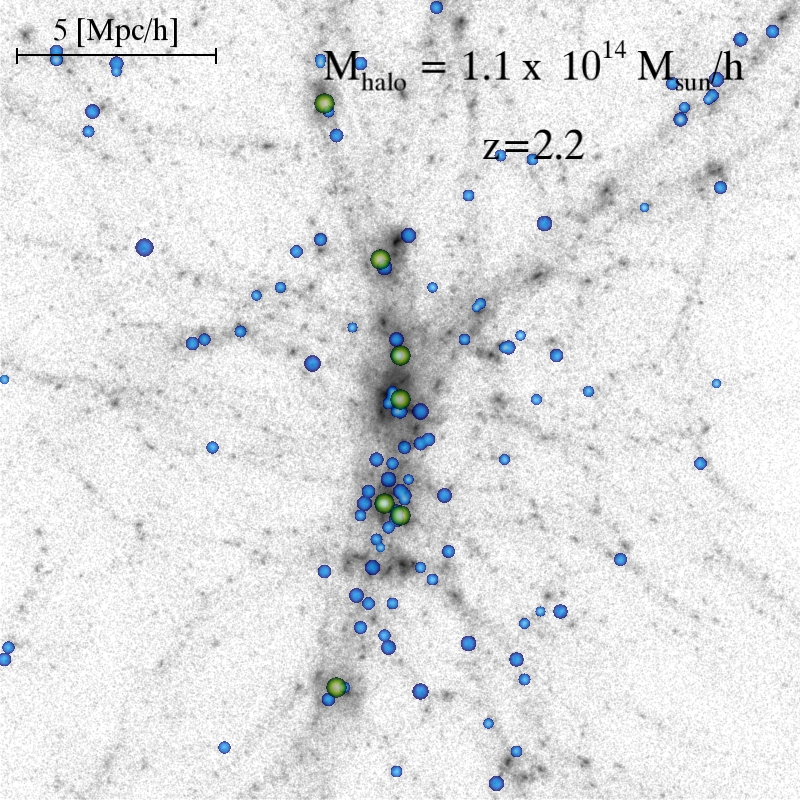}
\rotatebox{90}{\textcolor{red}{\Large \hspace{1.75cm} Quasars}}
\includegraphics[width=5.7cm]{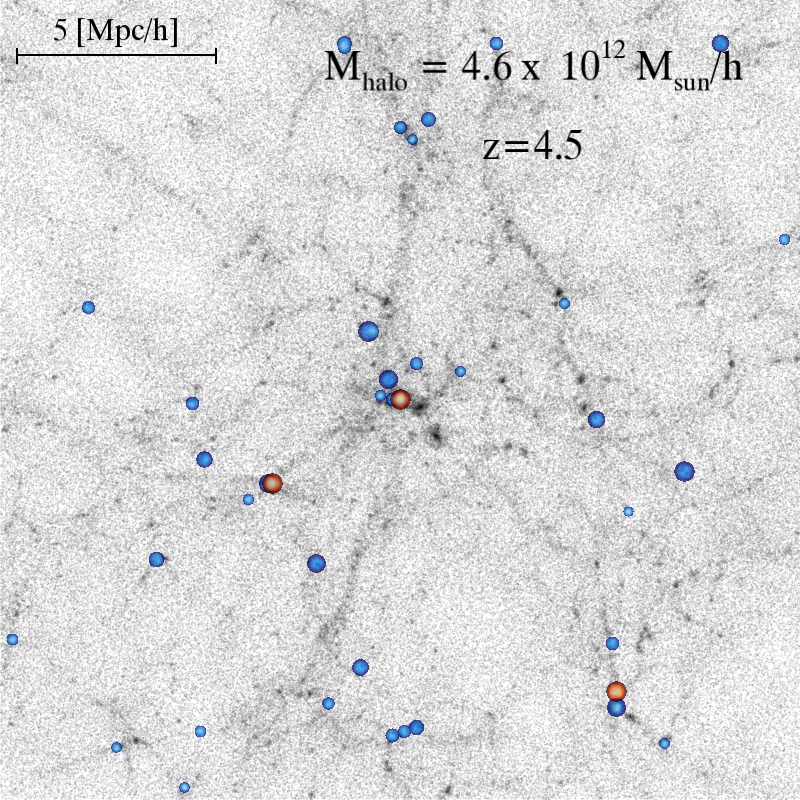}
\includegraphics[width=5.7cm]{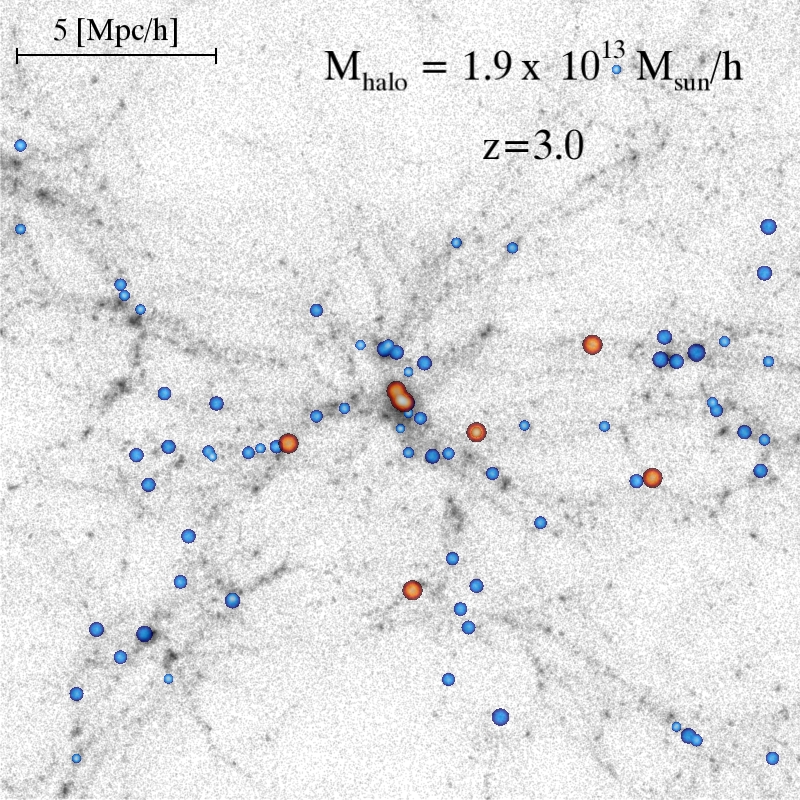}
\includegraphics[width=5.7cm]{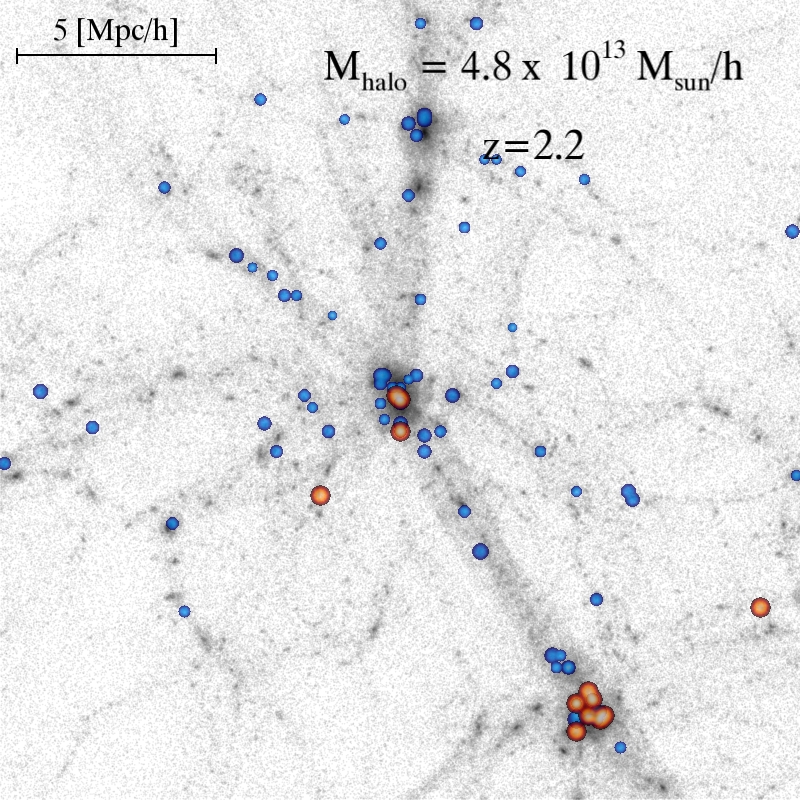}

\caption{The predicted spatial distribution of \lya\ emitters around typical radio galaxies (top) and quasars (bottom) at redshifts $4.5,
3.0$ and $2.2$ (from left to right). Each box shows a slice of $20 {\rm [Mpc/h]}$ on a side and $10 {\rm [Mpc/h]}$ depth. 
The DM density distribution is displayed in grayscale. Darker pixels
indicate a higher density of DM. 
Green circles show the position of the radio galaxies. Orange circles show the position of the quasars. Blue circles show the
spatial distribution of \lya\ emitters with $\llya > 10^{41} \lunits$}
\label{fig.images}
\end{figure*}

There is much observational evidence for the presence of outflows in \lya\ emitters at high redshifts \citep[e.g.][]
{giavalisco96, thuan97, kunth98, mas-hesse03,shapley03,kashikawa06,hu10,kornei10}. Hence, 
\citet{orsi12} assumed simple outflow geometries for the \lya\ photons to escape. 
For simplicity, two isothermal, spherically-symmetric models of galactic-scale outflows were adopted: 
an expanding thin shell and an expanding wind. Both are similar and their properties, such as 
expansion velocity, size and metallicity are directly proportional to the galaxy's predicted cold gas mass, circular velocity, 
half-mass 
radius and cold gas metallicity, respectively. In addition, the wind geometry displays a gas density profile of the form $
\rho(r) \propto r^{-2}$, where the normalisation depends on the mass-ejection rate predicted by the supernova feedback 
implemented in 
\galform. For simplicity, we will hereafter use the thin shell geometry. We 
have checked that our predictions are not sensitive to the choice of the outflow geometry. Since the \citet{orsi12}
model was developed using an earlier version of \galform\ \citep[described in][]{baugh05}, we re-calibrated the free parameters
controlling the relation between the half-mass radii of galaxies and the inner radius of the outflows for each galaxy. { The
new parameter values are found by matching the luminosity function of \lya\ emitters in the redshift range $0.2< z <6.6$
\citep[for details of the fitting procedure, see][]{orsi12}}
For starbursts, this relation is described by
\begin{equation}
 R_{\rm inner} = k(1+z)^\gamma \langle R_{1/2}\rangle,
\end{equation}
where $k = 0.2, \gamma = 1$ and $\langle R_{1/2}\rangle $ is the average of the half-mass radii of the disk and bulge 
components of the galaxies, weighted by their intrinsic \lya\ luminosity. For quiescent galaxies, 
$R_{\rm inner} = 2 \langle R_{1/2}\rangle$.

\subsection{Modelling of radio galaxies and quasars}

To model radio galaxies we use the AGN prescriptions described in \citet{fanidakis11}. The 
\citeauthor{fanidakis11} model
follows the mass accretion rate onto the BHs and the evolution of the BH mass, $M_{\mathrm{BH}}$, and spin, $a$, allowing the 
calculation of a variety of predictions related to the nature of AGN. In this model the evolution of BHs and their host galaxies 
is fully coupled: BHs grow during the different stages of the evolution of the host by accreting cold gas (merger/disk-instability 
driven accretion: starburst mode) and hot gas (diffuse halo cooling driven accretion: hot-halo mode) and by merging with other BHs.
 This builds up the mass and spin of the BH, and the resulting accretion power regulates the gas cooling and subsequent star 
formation in the galaxy. The resulting mass of the BH correlates with the mass of the galaxy bulge in agreement with the observations.
\citep[see][]{fanidakis11,fanidakis12}

The BH spin distribution depends strongly on how the gas in a given accretion episode accretes onto the BH. \citet{fanidakis11} 
assume that the accretion flow fragments due to self gravity into multiple accretion episodes \citep[chaotic accretion;][]
{king05,king08}. In this case, star formation in the vicinity of the BH can randomise the angular 
momentum of the gas, resulting in a succession of randomly aligned accretion disks around the BH. The end effect of this process is typically a BH 
with a low spin. High spin values occur only for the most massive BHs 
($M_{\mathrm{BH}}>10^8\mathrm{M}_{\odot}$), because the growth of these BHs is dominated by gas-poor BH-BH mergers which always result in fairly 
rapid spins of $a\sim 0.7-0.9$. Thus, in the chaotic accretion scenario there is a clear correlation of spin with BH mass and hence with host
galaxy bulge mass. Massive BHs form in the most massive DM halos, they are hosted by massive elliptical galaxies and have rapid 
spins, while lower mass BHs form in spiral galaxies and have much lower spins. 
{

The gas accreted during a starburst episode is converted into an accretion rate $\dot{M}$. The bolometric luminosity of the accretion
flow associated, $L_{\rm bol}$, will depend on the accretion rate in Eddington units $\dot{m} = \dot{M}/\dot{M}_{\rm Edd}$. If
$\dot{m} \ge 0.01$, then the thin disk solution of \citet{shakura73} is used 
\begin{equation}
L_{\rm bol} = \epsilon \dot{M}c^2,
\end{equation}
where $c$ is the speed of light, and $\epsilon$ is an adjustable parameter. Otherwise, the ADAF thick disk solution is adopted \citep{narayan94}
\begin{equation}
 L_{\rm bol, ADAF} = 0.44 \left(\frac{\dot{m}}{0.01}\right)\epsilon \dot{M}c^2.
\end{equation}
Finally, if the accretion is super-Eddington ($L_{\rm bol} \ge \eta L_{\rm Edd}$), then the bolometric luminosity is obtained as \citep{shakura73}
\begin{equation}
 L_{\rm bol}(\ge \eta L_{\rm Edd}) = \eta [1 + \ln (\dot{m}/\eta)]L_{\rm Edd},
\end{equation}
where $\eta$ and is an adjustable parameter of the model.
}

The mass, spin and mass accretion rate evolution are then coupled to the classic Blandford-Znajek jet model \citep{blandford77}. 
The jet power couples strongly to the accretion mode, as it most likely depends on the vertical (poloidal) magnetic field component 
$B_{\rm p}$ close to the BH horizon, $P_{\rm jet}\sim B_{\rm p}^2 M_{\rm BH}^2 a^2$. The accretion flow is assumed to form a 
geometrically thin disk for relatively high accretion rates \citep{shakura73}, switching at lower accretion rates to a geometrically 
thick disk in an advection dominated accretion flow \cite[ADAF;][]{narayan94}. The expression for the mechanical jet energy in each 
regime is then \citep{meier02}:
\begin{eqnarray}
P_{\rm{jet,ADAF}} & = & 2\times10^{45} M_9\left(\frac{\dot{m}}{0.01}\right)a^2 ~\rm{erg~s^{-1}},\\
\label{adaf_jet_luminosity}
P_{\rm{jet,TD}}&=&2.5\times10^{43} M_9^{1.1}
\left(\frac{\dot{m}}{0.01}\right)^{1.2}a^2 ~\mathrm{erg~s^{-1}},
\label{Jet_power}
\end{eqnarray}
where $M_9$ is the mass of the BH in units of $10^{9}{\rm M_{\odot}}$, $a$ is the spin of the BH. The collapse by two orders of magnitude in scale height of the flow during the transition from an ADAF to a thin disk
results in a similar drop in radio power. This already gives a dichotomy in radio properties which may explain some of the distinction between radio-loud
and radio quiet objects. Using these prescriptions for the accretion disk and jet, we calculate the optical and radio output from accreting BHs. The model
fits the luminosity function of radio-loud AGN remarkably well when low mass objects have lower jet powers than high mass objects. This is achieved
because the jet couples strongly to the BH spin and the lower mass BHs have lower spin than the most massive BHs. Overall, the model predictions for the 
AGN population can reproduce the diversity of nuclear activity seen in the local and high-$z$ Universe \citep{fanidakis11, fanidakis12}. 
The predictions for the radio-optical luminosities of FR-I, BLRG, Seyfert and LINER are in good agreement with the observed galaxy populations.

Based on the disk and jet luminosity the model calculates for every accreting BH, we define as a quasar any galaxy whose central BH produces a disk 
luminosity higher than $10^{46}~{\rm erg~s}^{-1}$. In contrast, a galaxy is defined as radio galaxy when its jet luminosity exceeds 
$P_{\rm 1.4GHz}=10^{23}~{\rm W~Hz}^{-1}$.  

\section{Results}
\label{sec.results}

\begin{figure}
\centering
\includegraphics[width=8cm]{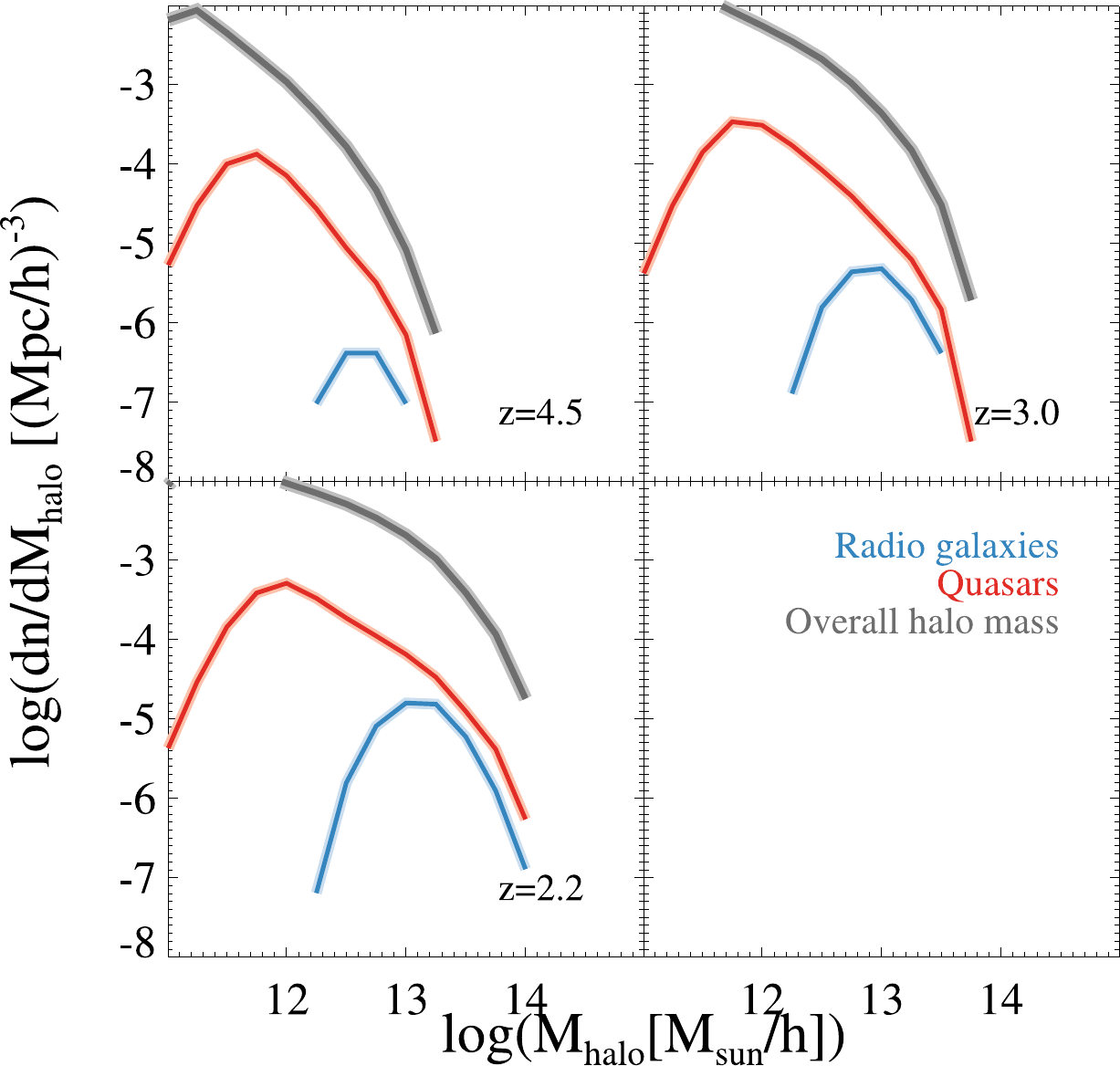}
\caption{The halo mass function of radio galaxies (blue) and quasars (red) in our model.
The grey line corresponds to the halo mass function of the overall galaxy population.
Panels correspond to redshifts $z= 4.5, 3.0$ and $z=2.2$ as labelled.}
\label{fig.mhalos}
\end{figure}

We now explore the predictions of \galform\ in the redshift range $2<z<6$. This is the redshift interval where
the bulk of the observational work about the environments around radio galaxies and quasars using \lya\ emitters has been carried out. 
Compiling statistical samples of \lya\ emitters is challenging at low redshifts $(z \lesssim 2)$, and has only been possible 
with the GALEX satellite  \citep[e.g.][]{deharveng08, cowie10,wold14}. 
At $z>6$, our model predicts that radio 
galaxies are very rare, such that it is not possible to robustly characterise the population.

Throughout this paper, we measure overdensities around radio galaxies and quasars in redshift space. This means that
the peculiar velocity of galaxies in the radial direction contributes to the observed redshift of galaxies, producing
a distortion of their derived comoving distances. To introduce this effect into our model, we take the comoving $z$-coordinate, $r_z$, to represent the line-of-sight direction and replace it by its value in redshift space $s$:

\begin{equation}
s = r_z + \frac{v_z}{aH(z)},
\label{eq.zspace}
\end{equation}
where $v_z$ is the peculiar velocity of the galaxy along the line-of-sight, $a$ is the expansion factor and 
$H(z)$ the Hubble parameter evaluated at redshift $z$.

Most of our predictions focus on four redshifts $z=2.2, 3.0, 4.5$ and $5.7$. These are the redshifts
at which \lya\ has been detected from the ground with negligible atmospheric contamination. In addition, $z=2.2$ is particularly 
important since it is also { the redshift at which a ground-based near-infrared (NIR) instrument can typically search for \ha\ emitters 
\citep[e.g.][]{geach08,koyama13}, although \ha\ searches have extended up to $z\sim 2.5$ \citep{cooke14}}

\begin{figure}
\centering
\includegraphics[width=8.5cm]{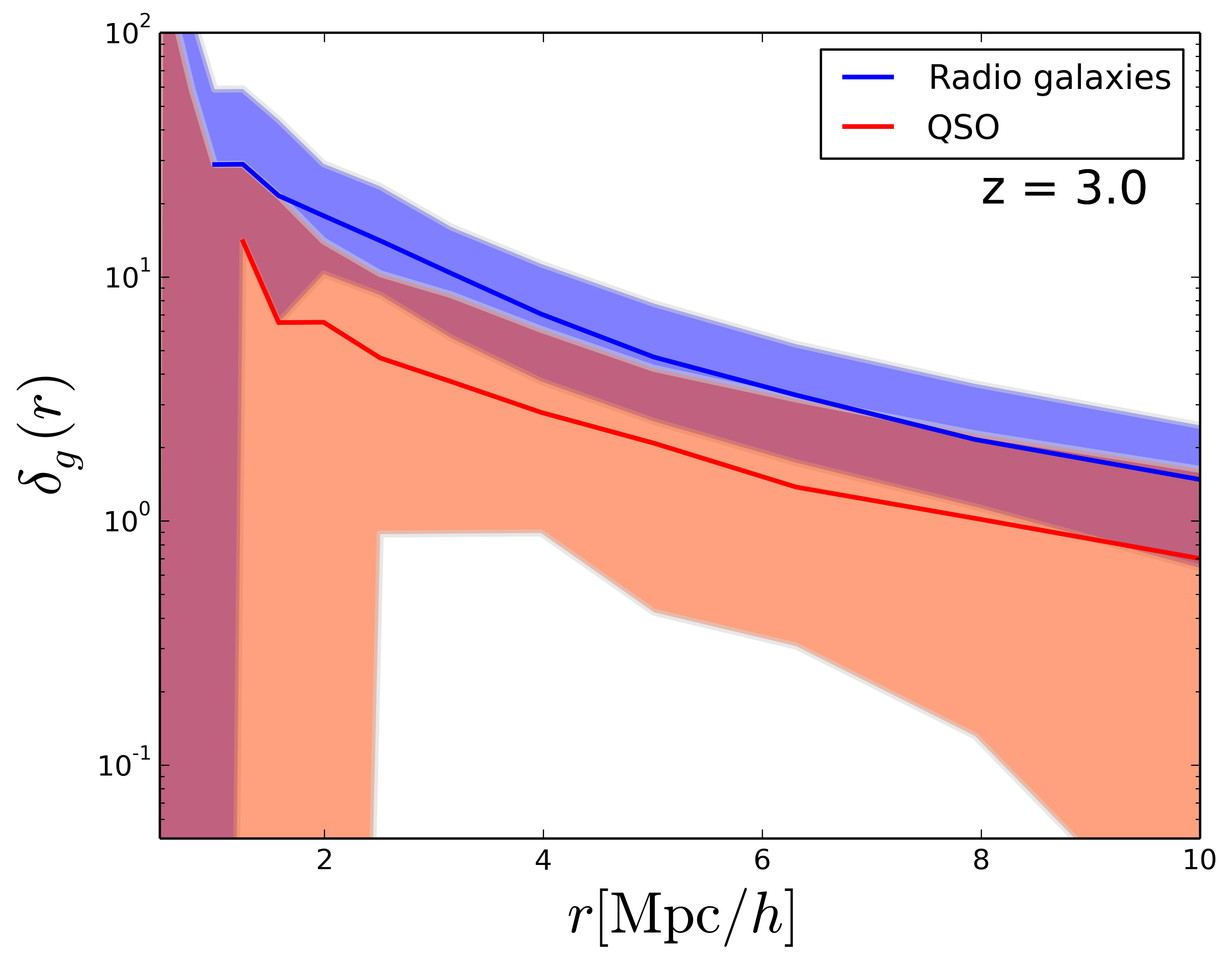}
\caption{The overdensity of \lya\ emitters, $\delta_g$, around radio galaxies and quasars at redshift $z=3$, 
as a function of the distance from the central object. 
The solid curve and the shaded regions show the median and 10-90 percentile of the overdensity distribution 
for a sample of \lya\ emitters with $\llya > 10^{41} \lunits$.}
\label{fig.delta}
\end{figure}

Fig. \ref{fig.images} shows the environment around selected radio galaxies and quasars within the redshift range 
discussed above, as predicted by \galform. Each image displays a cuboid of $20\times 20 \times 10 [{\rm Mpc}^3 h^{-3}]$, which illustrates 
the complicated filamentary structure of the DM, shown in grey. \lya\ emitters with luminosities 
$\logllya > 41$ are shown in blue. 

Radio galaxies and quasars trace different environments as a consequence of their different triggering mechanisms.
Quasar activity is triggered in gas-rich galaxies, typically associated with haloes of mass $10^{12}[{\rm M_{\odot}}/h]$ \citep{fanidakis13}.
For radio galaxies, we find that the brightest objects live in the most massive haloes because this is where 
the spin and mass of BHs are higher, but also the accretion rate is low enough to form an 
ADAF which gives powerful jets, as discussed in the previous section.

Fig. \ref{fig.mhalos} shows the mass function of haloes that host radio galaxies and quasars at 
different redshifts. Both types of AGN can be found in massive haloes. However, they represent 
a small subset of the population of massive haloes at any redshift. Furthermore, the fraction of 
haloes hosting a radio galaxy or quasar at a given halo mass is predicted 
to increase with redshift: At $z=2.2$, radio galaxies and quasars account for only 4.5 
per cent of the haloes with mass above $M_{\rm halo} > 10^{13} [{\rm M_{\odot}}/h]$. 
At $z=3.0$ this fraction increases to 5.2 per cent, and at $z=4.5$ to 12 per cent.

Quasars are significantly more abundant than radio galaxies at all redshifts and halo masses, 
and they also span a larger range of halo mass, peaking at $M_{\rm halo} \sim 10^{11.5}-10^{12} [{\rm M_{\odot}}/h]$. 
Radio galaxies, on the other hand, populate a small subset of the most massive haloes, peaking above 
$M_{\rm halo} \sim 10^{13} [{\rm M_{\odot}}/h]$ . 

The environments of these two types of AGN are mostly dominated by this fundamental 
difference in their halo mass distribution. Furthermore, baryonic processes that are important over these halo mass 
ranges can also have an impact on the properties of the galaxy 
population used to trace the environment and overdensities.

Observations and models have shown that star-forming galaxies tend to {\it avoid} the 
centres of massive structures \citep[e.g.][]{orsi10,contreras13}.
This is due to the star-formation quenching mechanisms that are expected to act in overdense regions.
The typical star-formation timescale that is traced by nebular emission is of the order $\sim 10 {\rm Myr}$. Hence, any baryonic 
process that results in an abrupt quenching of star-formation, such as AGN feedback, is expected to affect the 
line luminosities of galaxies, given its short timescale. Hence, the properties of emission-line galaxies (ELGs) around a massive 
structure will be related not only to the environment itself, but also to the depth (i.e. the limiting flux) of 
the galaxy sample used to trace such environments. Hereafter, we study the dependence of a number of properties on 
environment by splitting our ELG sample into 
"faint", i.e. those galaxies with line luminosity $L > 10^{41} \lunits$ which represent a deep survey, and "bright", 
with $L > 10^{42} \lunits$,
which represents a shallow survey. From the observational perspective, these represent complementary strategies. 
A deep and small survey can characterise the properties of galaxies within their host halo in a small volume, 
whereas a shallower and wider survey could be used to measure statistical properties, 
such as the galaxy clustering around these massive structures.

\subsection{The environment of overdense regions traced by radio galaxies and quasars}
\label{sec.clustering}

\begin{figure}
\centering
\includegraphics[width=8.5cm]{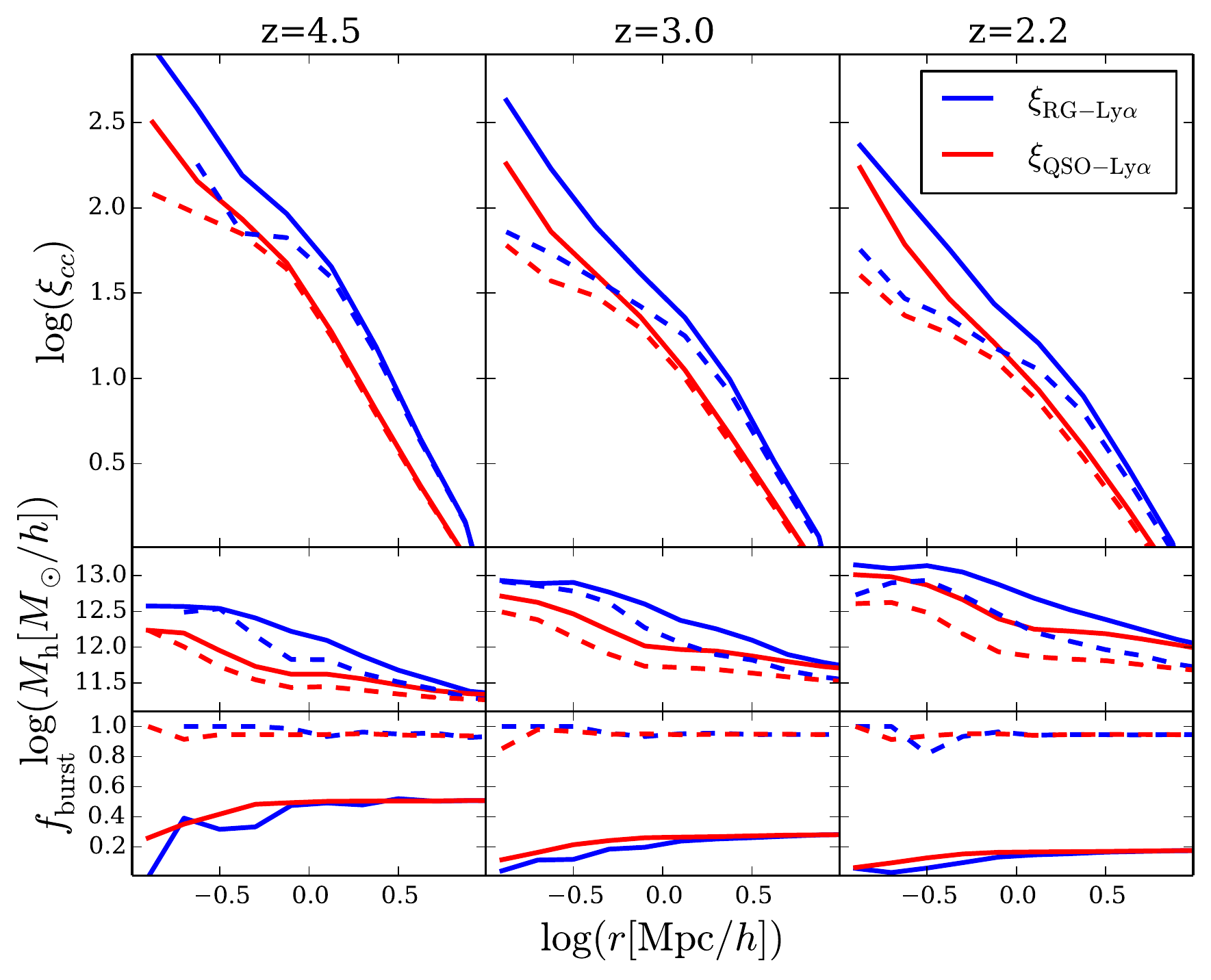}
\caption{  The clustering and properties of \lya\ emitters around radio galaxies (blue) and quasars (red) 
as a function of distance from the central object for redshifts $z=4.5, 3.0$ and $z=2.2$. Solid lines correspond to 
predictions for the faint sample 
of \lya\ emitters, whereas dashed lines correspond to the bright sample. 
(Top): the cross correlation function between radio galaxies, quasars, and 
\lya\ emitters.
(Middle): The mean halo mass of \lya\ emitters as a function of distance to the central object. (Bottom): 
The mean fraction of starburst galaxies as a function
of distance to the central object.}
\label{fig.cross}
\end{figure}

\begin{figure}
\centering
\includegraphics[width=8.5cm]{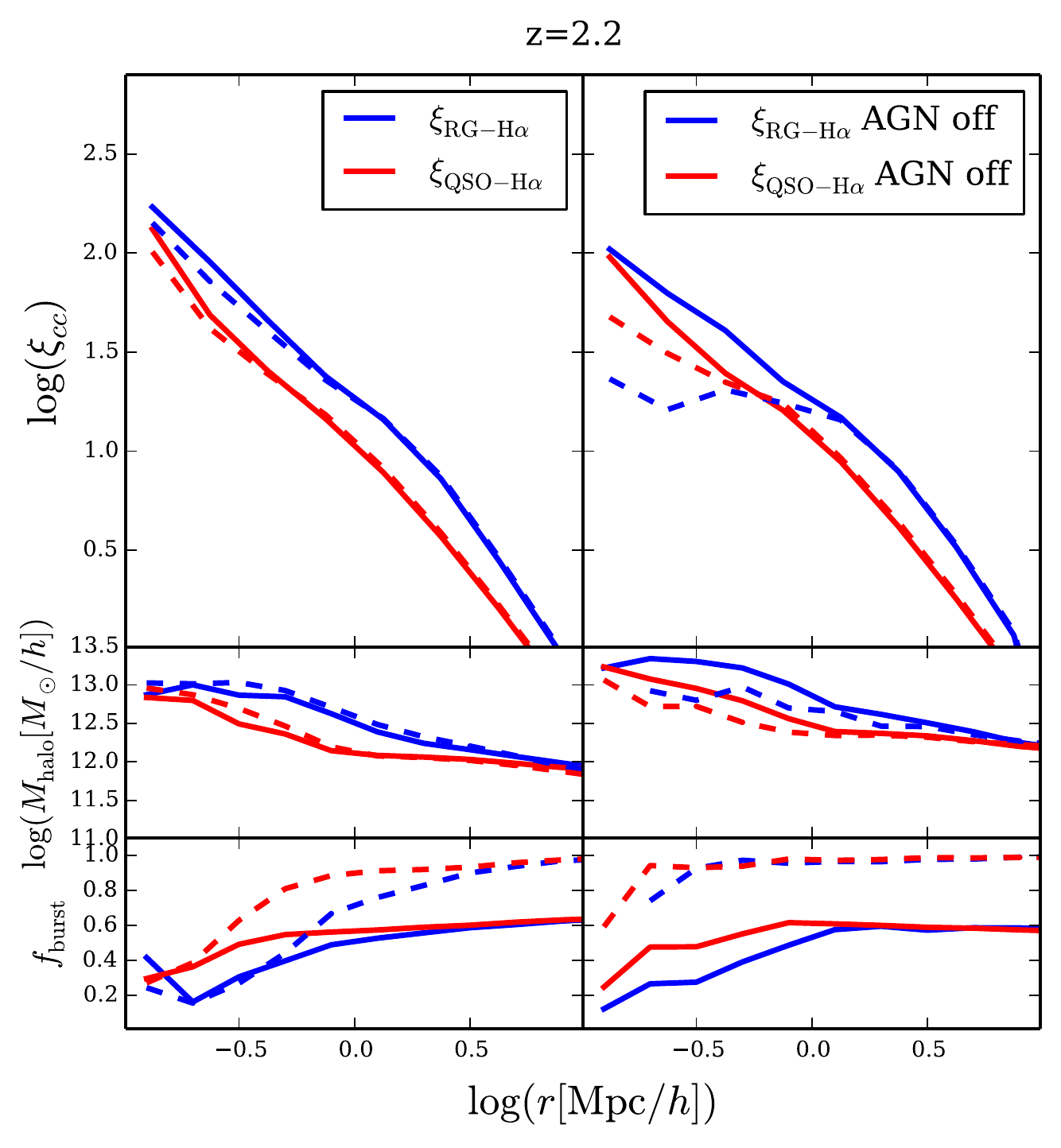}
\caption{Same as in Fig. \ref{fig.cross} but using \ha\ emitters at $z=2.2$. The left 
panel shows results for the fiducial model, whereas the right panels corresponds to the case in which AGN feedback has been turned off.}
\label{fig.cross_ha}
\end{figure}

We start by characterising the properties of overdense regions traced by radio galaxies and quasars.
The detection of these overdense regions at high redshifts typically requires long exposures and dedicated
observations via narrow-band imaging or spectroscopic follow-up. It is thus common for observational studies
to focus on one or only a handful of central objects 
\citep[e.g.][]{steidel05,venemans05,overzier06,venemans07,kuiper11,husband13,banados13,saito14,chiang15,adams15}. 
Their interpretation, from a galaxy formation perspective, is, thus, limited by cosmic variance and the lack of 
statistical samples. In addition, projection effects can smear out the overdensity significantly, even in narrow-band 
surveys \citep{chiang13}. 

In Fig. \ref{fig.delta} we illustrate the predicted overdensities of faint \lya\ emitters around radio galaxies 
and quasars at $z=3$ as a function of distance in redshift space. 
We define the galaxy overdensity as $\delta_g(r) = (n(r) - \bar{n})/\bar{n}$, where $n(r)$ is the number
density of galaxies within a sphere of radius $r$ around a central object, and $\bar{n}$ is 
the galaxy number density averaged over the simulation volume.

These predictions represent the ideal case in which no projection effects affect the measured overdensities.
Overall, the median overdensity around radio galaxies is predicted to be higher than that around quasars. 
This is consistent with the correlation between halo mass local overdensity. However, the scatter of 
$\delta_g$ around the median (shown in the plot as the 10 and 90 percentile range
of the distribution of overdensities around each radio galaxy and quasar in the model) is significant. For example, 
at distances below $\sim 1 \mpc$, the distribution of overdensites around radio galaxies and quasars can 
span several orders of magnitude due to the small number of galaxies at these distances.
At larger distances from the central object, overdensities of quasars display a larger scatter than around 
radio galaxies. This is likely to be caused by the larger range of halo masses that are hosts of quasars, 
as opposed to radio galaxies that are found in a much narrower halo mass range (see Fig. \ref{fig.mhalos}).

By studying a sample of overdensites around radio galaxies above $z>2$, \citet{venemans07} reported that 2 out of 8 
of their radio galaxies have environments that are consistent with being equivalent to the field. { 
Our model predicts that the fraction of radio galaxies hosted in average or underdense environments 
increases towards lower redshifts and lower host halo masses. Also, this fraction is significantly lower when measuring
the overdensity of galaxies with the faint samples of galaxies instead of the bright ones. At $z=2.2$, for instance, $20\%$ of radio galaxies
hosted by haloes with mass $M_{\rm halo}\sim 10^{12}$ are in environments with $\delta_{\rm gal} \leq 0$ of bright \lya\ emitters. This fraction is
reduced to $\sim 9\%$ when measuring $\delta_{\rm gal}$ with faint galaxies. In quasars, 
these fraction are generally higher, reaching up to  $35\%$ for the same halo masses. 
This population of underdense radio galaxies and quasars arise partly due to cosmic variance and, as shown 
in Section 4, due to the fact that most of the radio galaxies and quasars with $\delta_{\rm gal} \approx 0$ do not evolve to become 
massive clusters, but instead become average haloes at $z=0$.}


A better way to quantify the clustering is to compute 
the cross-correlation function between central quasars or radio galaxies and ELGs, $\xi_{cc}$. This is estimated as
\begin{equation}
\label{eq.xicc}
\xi_{cc}(r) = \frac{DD(r)}{N_{\rm c} n_{\rm gal} \Delta V(r)} -1,
\end{equation}
where $DD(r)$ is the total number of galaxies around central objects at a distance 
$r \pm \Delta r/2$, $N_c$ is the total number of central objects in the simulation box, $n_{\rm gal}$ is the mean
number density of galaxies, and $\Delta V(r)$ is the volume of a spherical shell of radius $r$ and
width $\Delta r$. This width corresponds to the bin size used to compute $\xi_{cc}$. { Eq. (\ref{eq.xicc}) is
suitable for computing the cross-correlation function because our simulation box is periodic, and thus, the pair counts 
are not affected by edge effects. As a result, there is no need to make use of estimators that rely on random sets of objects.}

Fig. \ref{fig.cross} shows the predicted cross-correlation function between radio galaxies and 
\lya\ emitters, and quasars and \lya\ emitters at three redshifts spanning $2.2\leq z \leq 4.5$. 
The amplitude of $\xi_{cc}$ is higher when using radio galaxies as the central objects than it is for quasars for any redshift, as expected.
An interesting feature arises at small scales (i.e. $r \lesssim 2 \mpc$), where $\xi_{cc}$ has a higher amplitude 
when computed using the sample of faint \lya\ emitters than when using the bright sample. This might seem counterintuitive 
at first, since we would expect brighter objects to be more clustered than faint ones.

\begin{figure}
{\Large \hspace{3cm} Ly$\alpha$ \hspace{2.75cm}     H$\alpha$}\par
\centering
\includegraphics[width=8.5cm]{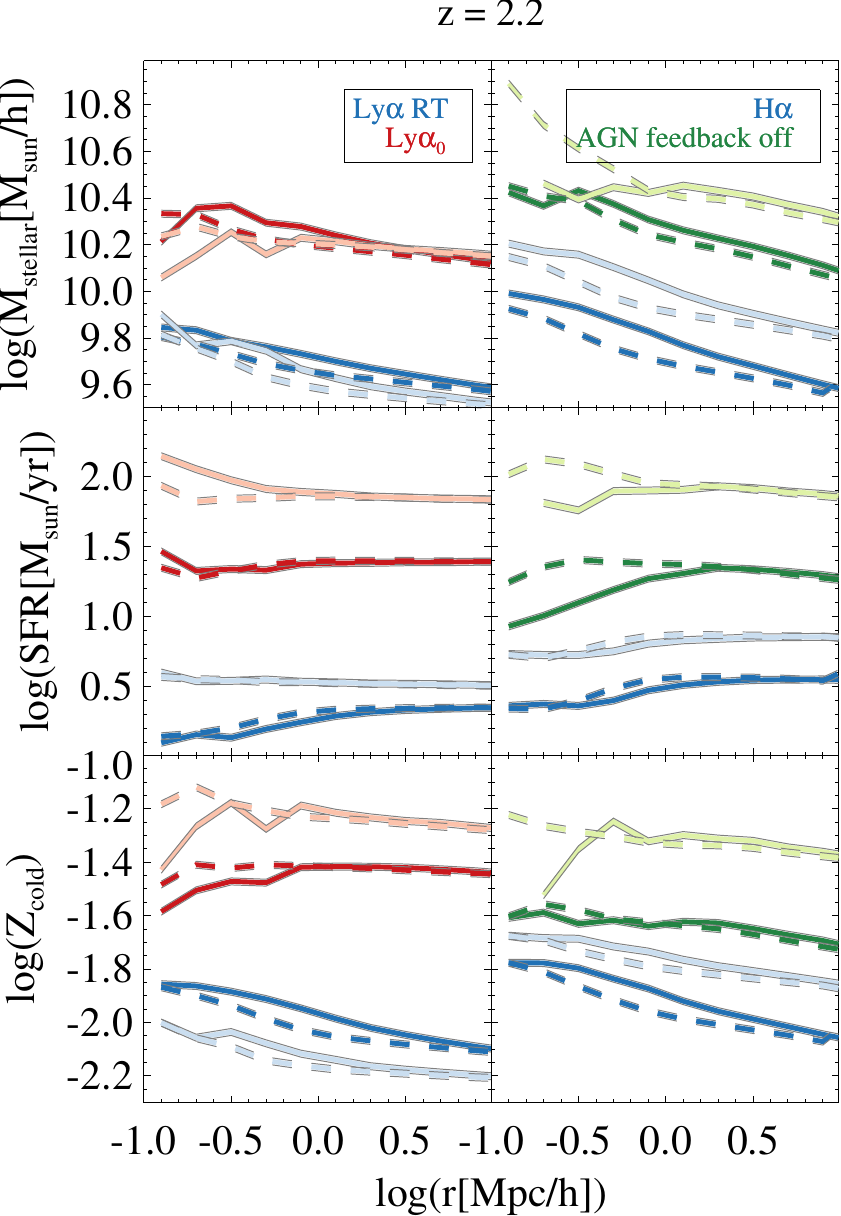}
\caption{ The mean stellar mass (top), SFR (middle) and gas metallicity (bottom) as a function of distance from radio galaxies and quasars at $z=2.2$. 
Solid curves are computed using environments around radio galaxies, whereas dashed curves represent the same around quasars.
The left column shows the properties of \lya\ emitters. Blue and red corresponds to the fiducial model and a variant without
radiative transfer effects, respectively. The right column shows the properties of \ha\ emitters. Here blue and green 
are the fiducial and a variant with no AGN feedback, respectively. In both columns, faint colours correspond to the sample
of bright galaxies, whereas darker colours correspond to the faint samples.}
\label{fig.env}
\end{figure}

To investigate the origin of this significant difference in the amplitude of 
clustering at small scales we look at the properties of the galaxies surrounding 
radio galaxies and quasars. The mean halo mass of \lya\ emitters as a function 
of distance is shown in the middle panels of Fig. \ref{fig.cross}. The mean halo 
mass of \lya\ emitters around radio galaxies is about 0.5 dex higher than that of \lya\ 
emitters around quasars. This is a consequence of the hierarchical growth of structures, 
in which massive haloes are surrounded by other massive haloes. In both
samples surrounding quasars and radio galaxies, faint \lya\ emitters are hosted by haloes 
that are slightly more massive than brighter ones. At 
separations of $r \gtrsim 10 \mpc$, the mean halo mass of both faint and bright galaxies 
tend to converge. This is translated into their cross-correlation functions having virtually 
the same amplitude.

{
\begin{table*}
\caption{Protocluster radius $r^* [{\rm Mpc}/h]$, median completeness/purity $C$, mean descendant halo mass $M(z=0)[M_{\odot}/h]$ and fraction of cluster descendants $f(M>10^{14}[M_{\odot}/h])$ 
of \lya\ samples around radio galaxies and quasars for different redshifts.}
\label{table.cp}
\begin{tabular}{@{}lccccccccc}
\hline
 Redshift & $r^*_{\rm radio}$ & $C_{\rm radio}$ & $\log M_{\rm radio}(z=0)$ & $f_{\rm radio}(M>10^{14})$ & $r^*_{\rm qso} $ & $C_{\rm qso}$ & $\log M_{\rm qso}(z=0)$ & $f_{\rm qso}(M>10^{14})$ \\
\hline
5.7 & 9.37 & 0.88 & 14.59 & 0.76 & 6.88 & 0.73 & 14.07 & 0.34\\
4.5 & 6.97 & 0.85 & 14.48 & 0.69 & 6.04 & 0.69 & 14.01 & 0.29\\
3.0 & 5.77 & 0.84 & 14.34 & 0.58 & 5.08 & 0.61 & 13.91 & 0.23\\
2.2 & 5.12 & 0.78 & 14.26 & 0.50 & 5.00 & 0.50 & 13.83 & 0.19\\
\hline
\end{tabular}
\end{table*}

}

To understand why faint \lya\ emitters are hosted by more massive haloes than bright \lya\ emitters are, we look at the relation between 
line luminosity and halo mass. Overall, both quantities are correlated. However, as shown in \citet{orsi12}, 
the correlation differs depending on whether galaxies are forming stars quiescently or in starbursts. In this
variant of \galform, quiescent galaxies form stars following a \citet{kennicutt83} IMF. { Starbursts, on 
the other hand, form stars with a top-heavy-like  IMF 
\citep[see][]{baugh05}. 
This IMF produces about 4 times more ionising photons for each star-formation episode than a ``normal'' IMF
\citep{ledelliou05}}. This implies that lower mass haloes that experience a starburst can be brighter in \lya\ 
luminosity than a more massive halo forming stars quiescently. The bottom panels
of Fig. \ref{fig.cross} show precisely this: the bright sample of \lya\ emitters consists almost 
entirely of starbursts, whereas the faint sample contains at most 40 per cent of starbursts. 

The difference in the fraction of starbursts between faint and bright samples of \lya\ emitters is also enhanced 
by the \lya\ radiative transfer model, which
assigns low escape fractions to quiescent galaxies, and higher ones to starbursts, as shown in \citet{orsi12}. 
Hence, the resulting cross-correlation functions shown in Fig. \ref{fig.cross} are the result of a combination 
of: AGN modeling (which results in radio galaxies populating more massive haloes than quasars),
the hierarchical clustering of haloes (which leads to more massive haloes surrounding radio galaxies than quasars), 
the choice of the IMF for quiescent galaxies and starbursts (which is responsible for most of the bright \lya\ emitters in lower 
mass haloes than quiescent \lya\ emitters), 
and finally the \lya\ radiative transfer model favouring the escape of \lya\ photons in starbursts. 

Unlike \lya\ emitters, \ha\ emitters are not subject to complex radiative transfer due to resonant scattering, 
and their attenuation by dust is smaller, making 
these galaxies excellent tracers of the instantaneous SFR \citep{kennicutt98b,calzetti13}. Apart from the radiative transfer 
effects, however, \ha\ emitters are essentially equivalent to \lya\ emitters in nature. 
Hence, by comparing the properties of the same environments traced by \lya\ and \ha\ emitters 
we can obtain a better picture of how environmental processes affect the galaxy properties.

We show the cross-correlation functions between radio galaxies, quasars and samples of faint and bright \ha\ emitters at 
$z=2.2$ in Fig. \ref{fig.cross_ha}. As with \lya\ emitters, the cross-correlation function involving radio galaxies 
has a higher amplitude than that with quasars as central objects. However, unlike with \lya\ emitters, there is little 
difference between the correlation functions when using the bright and faint samples of \ha\ emitters, 
even at small scales. The middle panels of Fig. \ref{fig.cross_ha} show that the mean halo mass for the two samples 
of galaxies is also very similar. Also, the fraction of starbursts, shown by the bottom panels of Fig. \ref{fig.cross_ha} 
shows that there is a similar fraction of starbursts at small scales, although the fraction of starbursts increases 
towards larger scales in the bright sample of \ha\ emitters. At large scales, there is only a
weak dependence between \ha\ luminosity and clustering amplitude, measured by the bias factor $b$ \citep{orsi08}.

In order to understand what is causing galaxies in the faint and bright samplea of \ha\ emitters to have similar halo 
masses and starburst fractions we explore the effect of an environmental process that plays a role only 
on small scales. Since central objects are active galaxies, we run a variant of the \galform\ model in which AGN feedback 
is switched off, leaving everything else the same. Interestingly, this variant
of the model mimics the suppression of the clustering amplitude on small scales that is also evident in the \lya\ emitter 
samples, although in this case the difference between the amplitude of the cross-correlations involving faint and 
bright samples is much stronger when the central objects are radio galaxies. This is consistent with the expectation 
that AGN feedback plays a stronger role in the environments of radio galaxies than in quasars. 
 { In the absence of AGN feedback there are two noticeable
changes in the properties of galaxies. On one hand, galaxies in massive haloes that would be quenched
by AGN feedback are now part of both the faint and the bright samples of H$\alpha$ emitters, thus 
increasing the average halo mass of both populations. 
On the other hand, the bulk of the quiescent galaxies do not reach luminosities above $10^{42} \lunits$, so the bright sample of galaxies consists almost entirely of starburts. Those galaxies, as discussed earlier, span a larger range of halo mass because of their higher production of ionising photons. Therefore, the bright sample has, on average, smaller halo masses than the average fainter ones. 
As a result, at small scales, the mean halo mass of faint H$\alpha$ emitters is higher than that of bright ones, resulting in a significant difference in their clustering amplitude."
}

The analysis above shows that although the properties of galaxies in protoclusters are fundamentally determined by the 
properties of their host haloes, the role of baryonic effects on the 
properties of galaxies in overdense regions is also predicted to be important.

To gain deeper insight into the effect of environment on galaxy properties we compute
how other galaxy properties change as a function of distance from their central objects.
In particular, we focus on the stellar mass $M_{\rm stellar}$, the star-formation rate SFR and the cold gas metallicity $Z_{\rm cold}$.
Fig. \ref{fig.env} shows the mean values of these properties as a function of distance from their central objects for both
\lya\ and \ha\ emitters at $z=2.2$. We also compute these quantities using two variants of our fiducial model, one in which
there is no \lya\ radiative transfer, and another in which AGN feedback is turned off.

The mean stellar mass of \lya\ emitters in the field is predicted to be about $10^{9.6} {[\rm M_{\odot}}/h]$, and 
increases about $0.2$ dex in the inner regions of the overdense regions. The stellar mass 
of galaxies in radio galaxy or quasar environments is predicted to be indistinguishable, even when comparing 
faint and bright samples. When \lya\ radiative transfer is switched off, the stellar mass is predicted to be about 0.6 dex
higher than in the previous case. For \ha\ emitters, bright galaxies are more massive than faint ones, and the effect
of the environment is more important. When AGN feedback is off, galaxies can grow in stellar mass reaching 
$\sim 10^{10.8}{[\rm M_{\odot}}/h]$ for small distances in bright \ha\ emitters.

The SFR of faint \lya\ emitters tends to decrease by up to about 0.4 dex in overdense environments. Bright \lya\ emitters, as expected, 
have higher SFRs, but they display no environmental dependence. When radiative transfer is disabled, the environmental effect on the 
faint sample is erased, and there is a slight increase of the SFR in bright \lya\ emitters. \ha\ emitters, on the other hand, also present 
a very small decrease of their SFR in overdense environments. Interestingly, the variants with AGN feedback turned off also present a small 
decrease in the SFR. { This occurs because in \galform\ when a galaxy becomes a satellite its hot halo is stripped. Such environmental 
effect reduces the total reservoir of gas to form stars, thus producing a quenching of star formation. Recently, \citet{peng15} found 
evidence that this effect is a primary mechanism for quenching star formation in a sample of local galaxies.}

Finally, gas metallicities tend to increase as a function of environment density for both \lya\ and \ha\ emitters. Interestingly, 
our model predicts that faint \lya\ emitters should have a higher gas metallicity than bright ones, but the opposite is true
for \ha\ emitter samples. This is due to \lya\ radiative transfer effects. When AGN feedback is off, galaxies are predicted
to have much higher gas metallicities. This is a natural consequence of them having also larger stellar masses and SFRs.

Despite the difference in halo masses hosting radio galaxies and quasars, and AGN feedback being less common in quasars than 
radio galaxies, there is very little difference in the environmental dependence of the properties studied here between
the two tracers. AGN feedback acts to shut down the SFR, thereby preventing the formation of more massive galaxies and 
also the chemical enrichment of their gas component. \lya\ radiative transfer favours the escape of \lya\ photons in 
galaxies with lower gas metallicities, stellar masses and SFRs than is the case for \ha\ emitters.

\begin{figure}
\centering
\includegraphics[width=8.5cm]{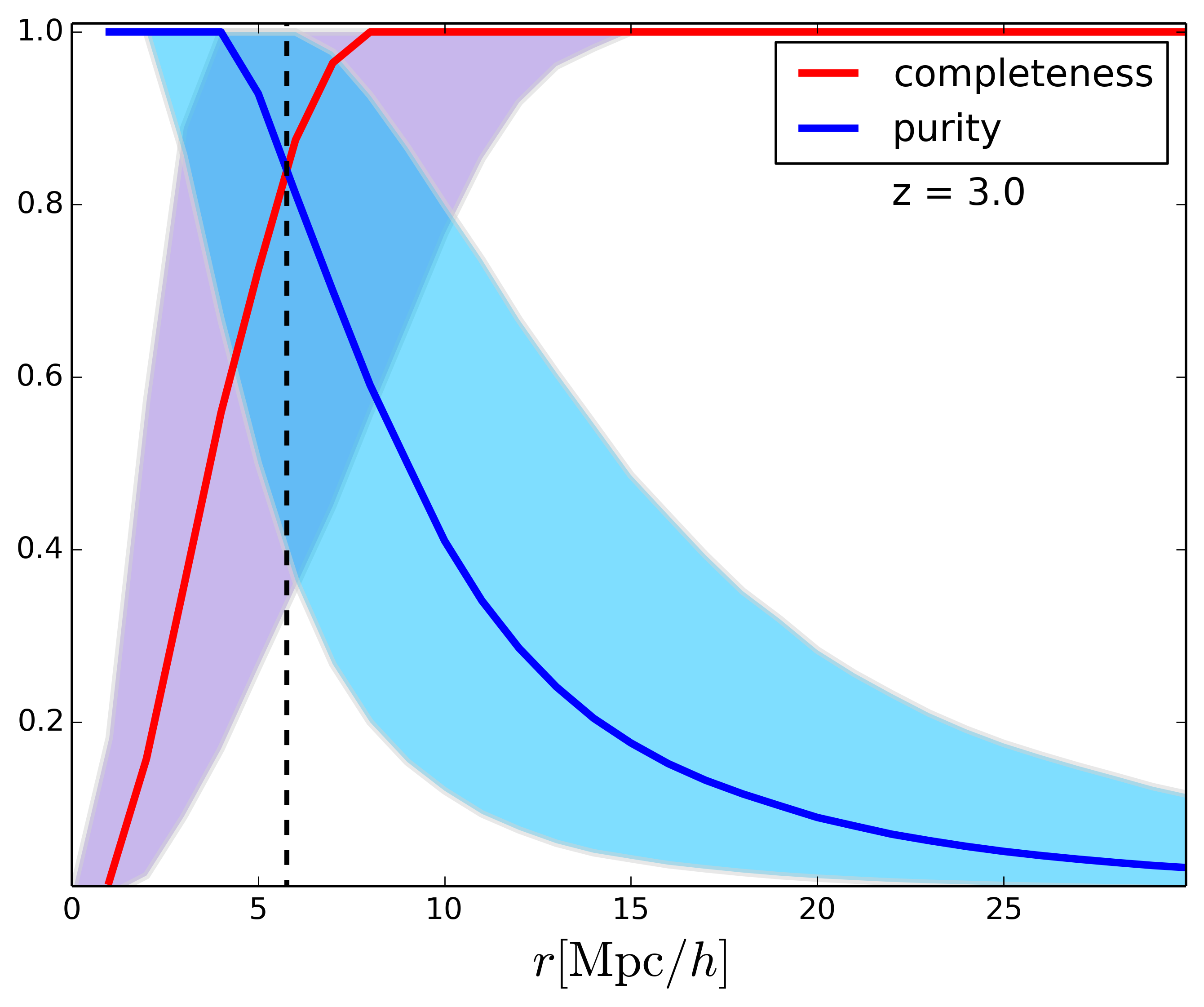}
\caption{The protocluster completeness (red) and purity (blue) of the sample of faint \lya\ emitters around radio galaxies at $z=3.0$. The solid lines show 
the median completeness and purity. The shaded regions show the 10-90 percentile of the distribution 
of completeness and purity of \lya\ emitters at a given distance from a central radio galaxy. The vertical dashed line
corresponds to the distance at which the completeness and purity are equal.}
\label{fig.cp}
\end{figure}

\begin{figure*}
\raggedright
{\Large \hspace{3.5cm} Radio galaxies \hspace{7cm}     Quasars}\par
\centering
\includegraphics[width=8.8cm]{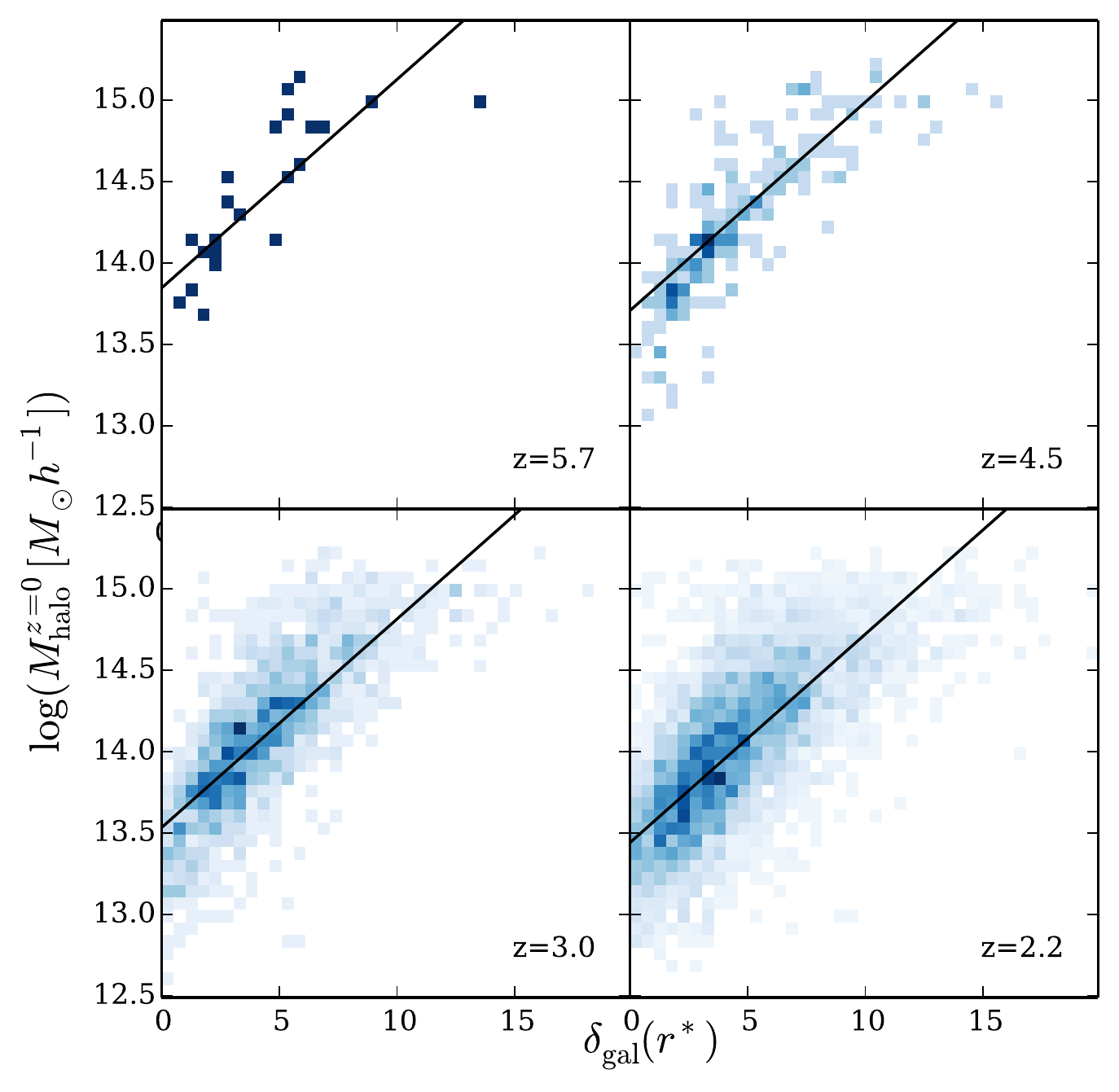}
\includegraphics[width=8.8cm]{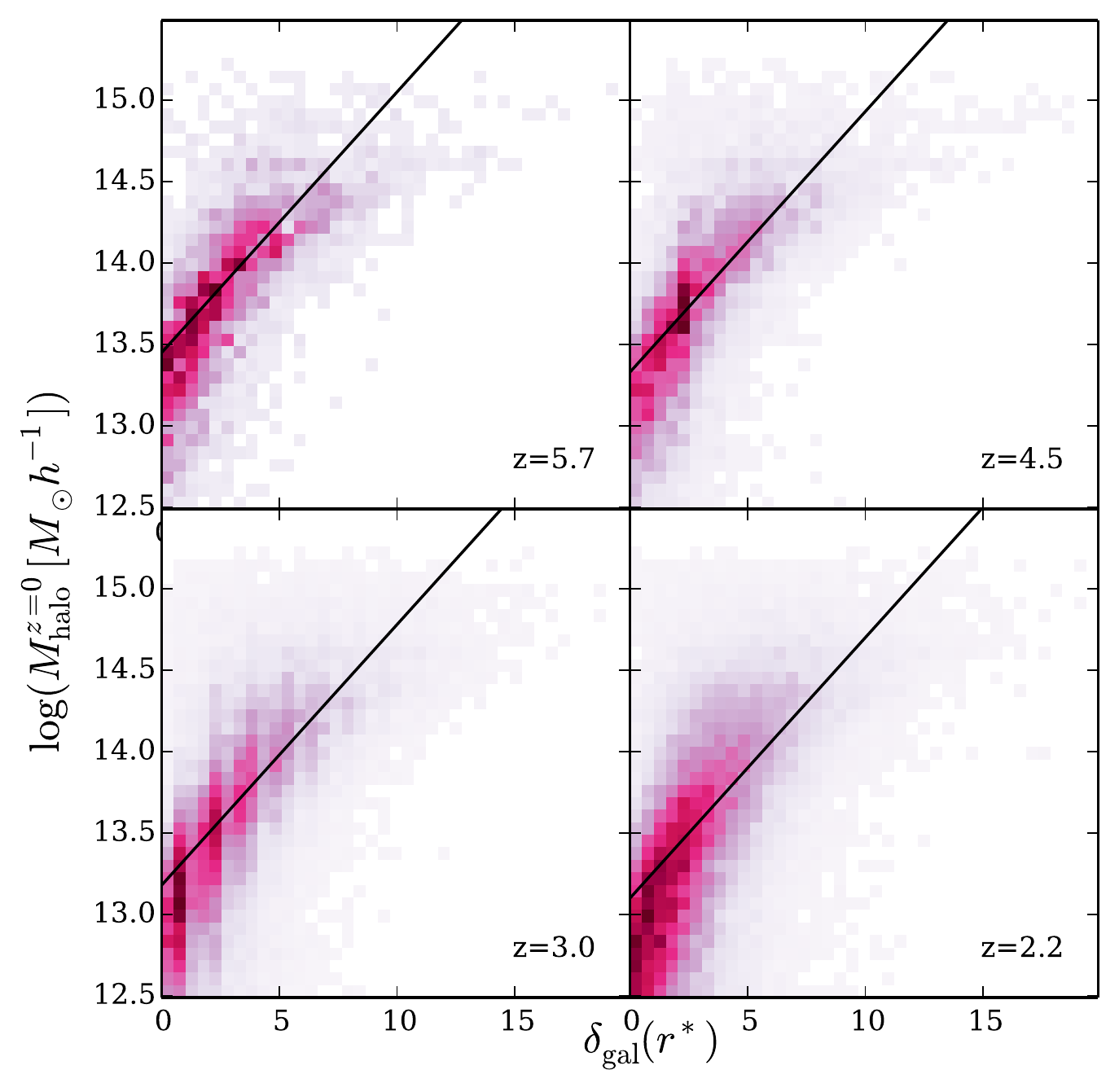}
\caption{The relation between protocluster galaxy overdensity, $\delta_g$, { measured with the faint sample of \lya\ emitters} within $r^*$ and their 
$z=0$ descendant halo mass $M^{z=0}_{\rm halo}$. The left panels show this relation for protoclusters traced by radio galaxies at different redshifts, 
and the right panels with quasars. Each panel correspond to a different protocluster redshift, ranging from $z=4.5$ down to $z=2.2$ as labelled. 
}
 \label{fig.delta_m0}
\end{figure*}

\subsection{Protoclusters and their descendants}
\label{sec.proto}

We now study how the overdense regions discussed before are related to progenitors of massive structures at the present day.
\galform\ can track the progenitors of any present-day halo by retrieving its merger history. This is useful for determining
which galaxies surrounding a central object at high redshift actually form part of a massive cluster at late times. We
determine the progenitors associated to a { protocluster-like structure} as those galaxies that surround a 
central object (radio galaxy or quasar) that at $z=0$ { are within the same (Friends-of-friends) massive DM halo}. { Furthermore, we
identify protoclusters as those massive structures that become part of a DM halo with mass $M_{\rm halo} \ge 10^{14}[M_{\odot}/h]$. As 
we discuss below, a significant fraction of radio galaxy and quasar environments do not form a massive cluster by $z=0$. Nevertheless, 
in the following we apply the same analysis to all environments around these central objects.}

A practical problem in observations is to determine the size of a protocluster. Since we can identify the progenitors of 
a massive halo, here we devise a method to determine the average size of { overdensities} around both types of central objects. 
We take the simple approach of considering
protocluster members as those galaxies that lie within a sphere of a given radius around the central object. If that radius 
coincides with the most distant progenitor, then the sphere will contain all true progenitors, but also a significant fraction 
of galaxies that are not progenitors of the same halo. Hence, to find an appropriate radius, we measure the completeness 
of a protocluster at a radius $r$, defined as the fraction of progenitors enclosed within that radius $r$. 
Likewise, we measure the purity of the sample at a radius $r$, defined as the ratio of the number of galaxies that are progenitors to the total
number of galaxies enclosed within $r$. { In the following, for simplicity, we will restrict the analysis of overdensities and progenitors
to those determined using the sample of faint \lya\ emitters.}

Fig. \ref{fig.cp} illustrates this by showing the distribution of completeness and purity as a function of distance 
from radio galaxies at $z=3$. As expected, at small distances from the central objects the purity is at its highest, 
and the completeness is low. At a given radius $r^*$, both quantities intersect, typically at high values. We call $r^*$ 
the typical protocluster radius at that redshift, and $C$ the corresponding completeness and purity at that radius.

Table \ref{table.cp} shows $r^*$ and the completeness and purity at that radius for radio galaxies and quasars at different 
redshifts. Overall, $r^*$ decreases towards lower redshifts. For radio galaxies, for instance, $r^*$ goes from 9.37 to 5.12 $\mpc$ 
for redshifts $z=5.7$ to $z=2.2$. { Radio galaxies at lower redshifts are, on average, hosted by more massive haloes than those at higher redshifts, 
and this is effectively translated into larger structures (in physical coordinates) at low redshifts than at high redshifts. 
However, this also implies a decrease of the radius of protoclusters towards lower redshifts when sizes are expressed in comoving units. }

Also, the size of { the proto-structures} traced by radio galaxies have completeness and purities around $\sim 0.8$ at all redshifts. 
Quasars, on the other hand, have lower values of completeness and purity, ranging from 0.5 at $z=2.2$ to 0.7 at $z=5.7$. 
This suggests that the $r^*$ values shown in Table \ref{table.cp} are reasonable for radio galaxies, but do a poorer job 
of representing the size of { proto-structures} traced by quasars. This is likely to be related to the large range of halo masses that 
host quasars, which is translated into a large variety of sizes and environments traced by quasars, unlike radio galaxies 
that are significantly more restricted in host halo mass (see Fig. \ref{fig.mhalos}). This also results in 
our model predicting radio galaxies to trace more massive descendant halo masses compared to quasar descendants. 
{ The fraction of protoclusters (i.e. those structures with $z=0$ halo mass $M_{\rm halo} \ge 10^{14}{\rm [M_{\odot}}/h]$) 
traced by radio galaxies and quasars differs significantly. At $z=2.2$, half of radio galaxies are predicted to trace protoclusters. On the other hand,
only $19\%$ of quasars pinpoint the location of protoclusters at this redshift. These fractions grow considerably towards higher redshifts, since both radio
galaxies and quasars trace increasingly higher and rarer peaks of the matter density distribution of the Universe towards high redshifts.}

{ Since the overdensity of galaxies is a biased tracer of the matter content, we explore how the 
overdensitiy within protoclusters correlate with their descendant $z=0$ halo mass.} To do this we compare the overdensity of 
galaxies measured within $r^*$ with the $z=0$ descendant mass for protoclusters traced by radio galaxies and by quasars. 
This relation is shown in Fig. 
\ref{fig.delta_m0}.
Overall, protoclusters traced by radio galaxies become haloes at $z=0$ with masses ranging over 
$10^{13.5} \lesssim M_{\rm halo}^{z=0} {\rm [M_{\odot}}/h] \lesssim 10^{15}$. Quasar descendants 
also reach those high halo masses, but a significant fraction have masses below $10^{13} {\rm [M_{\odot}}/h]$ 
\citep[for a discussion on quasar descendants see][]{fanidakis13}. The overdensity of galaxies within $r^*$ correlates 
well with the descendant halo masses for both radio galaxies and quasar descendants. 
This shows that our choice of radius $r^*$ to
determine the average extent of protoclusters at a given redshift is a resonable one.

The correlation between the galaxy overdensity and the halo descendant mass shown in Fig. \ref{fig.delta_m0} can be described with a simple linear 
form:
\begin{eqnarray}
 \label{eq.mrad_0}
 \log(M_{\rm rad}^{z=0}[{\rm M_{\odot}/h}])  & = & 13.07 + 0.11(1+z) + 0.12\delta_{\rm gal}(r^*), \\
 \log(M_{\rm QSO}^{z=0}[{\rm M_{\odot}/h}])  & = & 12.78 + 0.10(1+z) + 0.16\delta_{\rm gal}(r^*), 
 \label{eq.mqso_0}
 \end{eqnarray}
where $M^{z=0}$ is the descendant halo mass in units of ${\rm [M_{\odot}}/h]$. Although the constants in Eqs. (\ref{eq.mrad_0}) and (\ref{eq.mqso_0}) 
have similar values, their difference reflects the smaller descendant masses of quasars with respect to radio galaxies. Also, the scatter of the descendant
mass as a function of $\delta_{\rm gal}$ is larger in quasars than in radio galaxies, reflecting the broader diversity of environments that are traced
by quasars. Despite the apparent broad scatter around the scaling relations, we have checked that, for radio galaxies, 68 percent of the distribution
is within 0.2 dex of the relation given by Eq. (\ref{eq.mrad_0}). For quasars, this remains the same except at $z=2.2$, where 68 percent of the
distribution is within 0.6 dex of the scaling relation of Eq. (\ref{eq.mqso_0}).

Observational samples could make use of Eqs. (\ref{eq.mrad_0}) and (\ref{eq.mqso_0}) using $r^*$ predicted by \galform\ to obtain an approximate value of the $z=0$ 
descendant that a protocluster is expected to evolve into. { The scaling relations obtained when using the bright sample of Ly$\alpha$ emitters 
is consistent with the ones obtained for the faint sample within the $1-\sigma$ region, making Eqs.(\ref{eq.mrad_0}) and (\ref{eq.mqso_0}) suitable for 
samples of galaxies limited by these two luminosity ranges.}

{ The above analysis results in simple scaling relations that could be easily applied to observational data.} 
However, observations generally give projected measurements of the overdensity of galaxies, and these
are expected to differ significantly with respect to the spherically averaged overdensities 
\citep[e.g.][]{muldrew12,shattow13,chiang13}.
We can use our model to determine what overdensity values typical instruments would measure at different redshits. To 
do this, we project positions along the redshift space $s$ coordinate (see Eq. \ref{eq.zspace}), and convert the 
spectral resolution of an instrument into a comoving distance probed along the line-of-sight. 
We then compute the overdensity around a given protocluster by using a cylinder of radius $r^*$ and 
depth given by the spectral resolution of the instrument. In the case 
of very high resolution instruments, $r^*$ can be larger than the comoving distance range probed 
by the spectral resolution. Such cylinders would then probe only a fraction of the galaxies in a protocluster overdensity. 
To include all galaxies within the protocluster radius, we stack several cylinders of equal depth and decreasing radius around both sides of a central 
cylinder of radius $r^*$ to probe a volume that approximates that of a sphere of the same radius.

\begin{figure}
 \centering
 \includegraphics[width=8cm,angle=90]{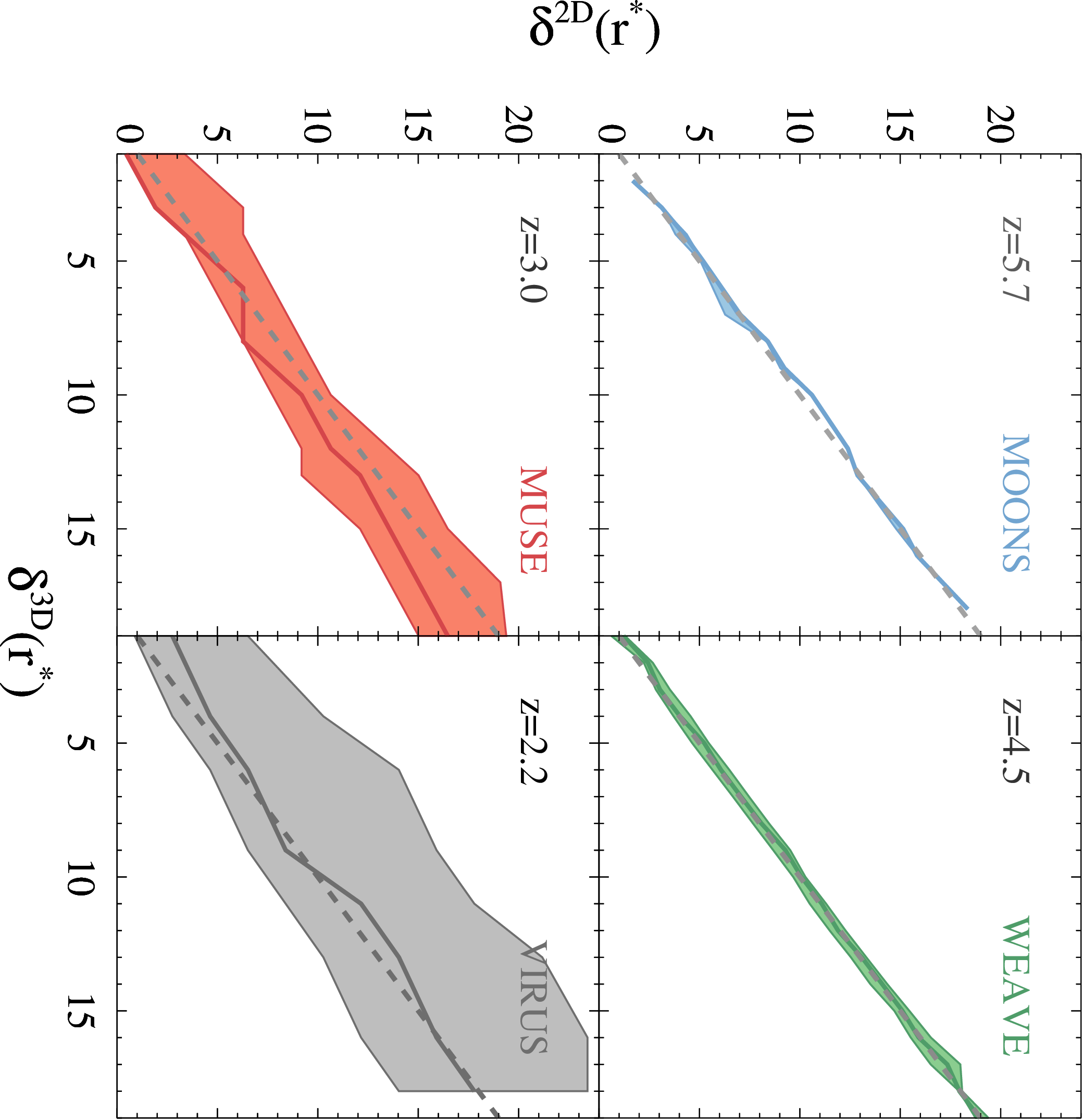}
 \caption{ The overdensity of protoclusters traced by radio galaxies and faint \lya\ emitters at different redshifts. 
 The panels compare the overdensity measured within a sphere, $\delta^{3D}$ 
 and that measured with projected distances along the line of sight within a cylinder, $\delta^{2D}$. The distance probed
 along the line of sight varies depending on the spectral resolution of the instrument simulated, shown in the legend of 
 each panel.}
 \label{fig.dproj}
\end{figure}

Fig. \ref{fig.dproj} shows a direct comparison between the overdensity computed within a sphere, $\delta^{3D}$ and that
within cylinders by projecting distances along the line-of-sight, $\delta^{2D}$, both measured at $r^*$ at their respective redshifts according to Table \ref{table.cp}. To illustrate the effect of different spectral resolutions, we choose four instruments, one for each redshift spanning $5.7\leq z \leq 2.2$. 

MOONS \citep{cirasuolo14} is a future multi-object spectrograph proposed for the VLT that, given its spectral coverage, can be used to find \lya\ emitters at $z=5.7$ and beyond. Its wavelength resolution, $\Delta \lambda = 0.9$ \AA{} at the \lya\ observed wavelength, translates into a 
line-of-sight comoving resolution at $z=5.7$ of $\Delta r_z = 0.24 \mpc$.  Fig. \ref{fig.dproj} shows that our model predicts a 
remarkably accurate projected overdensity compared to the real, spherical one. Similarly, WEAVE \citep{dalton12}, another future multi-object spectrograph at the 4.2 meter William Herschel Telescope on La Palma can be used to search for \lya\ emitters
at $z=4.5$. Its spectral resolution at the \lya\ wavelength translates into a comoving resolution of $\Delta r_z = 0.12 \mpc$. This allows to 
obtain very accurate projected overdensities, similarly to MOONS at $z=5.7$. 

At lower redshifts, a given spectral resolution is translated into larger comoving distance ranges, meaning larger projection effects. At $z=3.0$, for instance, we predict the projected overdensities 
that could be obtained with MUSE, a panoramic integral field unit (IFU) mounted on VLT \citep{bacon10}.  Its wavelength resolution at $z=3.0$ translates into 
$\Delta r_z = 1.33 \mpc$. The median value of $\delta^{2D}$ falls very close to the exact, spherical overdensity 
$\delta^{3D}$, although the scatter around is larger than in the previous cases due to 
the larger comoving distance resolution. Finally, for $z=2.2$ we illustrate the results of using VIRUS, an array of IFUs mounted on the Hobby-Eberly Telescope to carry our the HETDEX cosmological survey \citep{hill08}. The spectral resolution for \lya\ emitters at $z=2.2$ translates into
a comoving resolution of $\Delta r_z = 5.08 \mpc$, which is comparable to the typical size of protoclusters at this redshift. The scatter is, as expected, larger than in the previous cases. It is worth noting that both VIRUS and MUSE can obtain projected overdensities with much better accuracies for higher redshifts than shown here. 

On the other hand, narrow band filters with a typical width of $\sim 100$\AA{} can deliver projected overdensities with resolutions along the line of sight of about $50-100 \mpc$ over the redshift range shown here \citep[e.g.][]{venemans07,kuiper11,saito14}. Therefore, such projected overdensities are not suitable for characterising the scales of protoclusters as shown here. However, given the wider areas that are accesible with a photometric survey, such studies could characterise the clustering around radio galaxies and quasars over scales beyond the protocluster typical radii to put constraints on the DM halo masses hosting these two central objects.

\section{Conclusions}
\label{sec.conclusions}

In this paper we have presented predictions for the properties of overdense regions around radio galaxies and quasars at high redshifts using a 
model that includes a physical treatment for quasars, radio galaxies and star-forming emission-line galaxies. 

Most of our results here follow from a key prediction of our model, namely that, at a given redshift, 
radio galaxies are hosted on average by DM 
haloes that are significantly more massive than quasars. Furthermore, most quasars are hosted by DM haloes 
of masses $\sim 10^{12} {\rm [M_{\odot}/h]}$ almost irrespective of redshift, whereas radio galaxies populate the most 
massive haloes present at any redshift. This crucial distinction is translated into their clustering properties being different, and 
the impact of AGN feedback being stronger in radio galaxy environments. 

This paper discusses two problems. First, we study the properties of the overdense regions
traced by radio galaxies and quasars at $z>2$, measuring the environment with samples 
of emission-line galaxies and exploring how the physics of baryons impacts the galaxies
in these environments at high redshift. Second, we identify the progenitors of the descendant halo 
that is traced by radio galaxies and quasars to define protoclusters and link their properties to 
their descendant haloes.

The analysis of individual overdense regions at high redshifts is subject to significant cosmic variance that
limits their interpretation in terms of general galaxy formation physics. This explains, for instance, 
why some authors have found peculiar environments that are average or underdense around both radio galaxies and 
quasars \citep{venemans07,husband13,banados13}.

The cross-correlation functions $\xi_{cc}$  between overdensity tracers (radio galaxies and quasars) and emission-line
galaxies at high redshifts offer different information on small and large scales. At large scales ($r \gtrsim 10 \mpc$) the
amplitude of $\xi_{cc}$ is larger when the central objects are radio galaxies, because these are hosted by more massive
haloes. More interestingly, if we split the sample of emission-line galaxies into faint and bright galaxies, we find that 
the clustering on small scales is very different for both. 
For \lya\ emitters, radiative transfer effects and the higher production rate of ionising photons in starbursts invoked in
\galform\ makes faint \lya\ emitters have a higher $\xi_{cc}$ than bright ones. In the case of \ha\ emitters, which
are not affected by radiative transfer effects, the difference between faint and bright samples is insignificant. 
In this case, we find that AGN feedback prevents starbursts from dominating the galaxy abundance at small separations.
This effect is stronger in radio galaxies. 

The next generation of large redshift surveys will be able to characterise the environments of overdense regions with
unprecedented detail. In particular, a number of these will rely on emission-line galaxies to map the matter distribution. 
Spectroscopic surveys such as HETDEX \citep{hill08} and DESI \citep{levi13} have sufficient spectral resolution to probe the 
scales on which our model predicts that baryonic effects are noticeable on small scales. Other multi narrow-band surveys 
such as J-PAS \citep{benitez14} and PAU \citep{castander12} will provide large samples at $z\gtrsim 2$, and could measure
the amplitude of $\xi_{cc}$ on larger scales, thus allowing tight constraints to put on the halo masses hosting 
radio galaxies and quasars.

\lya\ radiative transfer and AGN feedback also have an impact on the physical properties of emission-line galaxies 
in overdense regions. Overall, the stellar mass, star-formation rate and gas metallicity of \lya\ emitters have 
a small environmental dependence, which is stronger in the absence of \lya\ radiative transfer. AGN feedback
produces average values of these properties in overdense regions that are different in radio galaxies than quasars, 
but the difference is too small to be probed observationally.

The progenitors of a present-day cluster allow us to define in the model an average protocluster radius $r^*$ in overdensities
traced by radio galaxies and quasars, by determining the distance at which the average completeness and purity of the 
predicted protoclusters coincide. This simple definition allows us to correlate the galaxy overdensity within
the protoclusters with the halo descendant mass at $z=0$. We compute scaling relations between these two quantities
that should provide with a good approximation the halo descendant mass of observed protoclusters at high redshifts.

Current and planned high resolution multi object spectrographs and IFUs are predicted to be able to measure 
projected overdensities that are very close to the ones computed in 3D, 
thus making it feasible to explore the predictions of this work with regard to observational data.


The current generation of instrumentation with resolutions above $R\sim 3000$ should be able to resolve the 
inner structure of protoclusters, where the strongest departures of the galaxy properties with respect to the field
are predicted to occur. On the other hand, wide-area narrow-band surveys should be able to provide large statistical 
samples to probe the clustering properties of protoclusters traced by quasars and radio galaxies. 

\section*{Acknowledgements}
We would like to thank Nelson Padilla and Andrea Maccio for encouraging discussions about this project. 
AO acknowledges support from Fundaci\'on ARAID and FONDECYT project 3120181. CMB acknowledges support 
from STFC Consolidated grant ST/L00075X/1.
Part of the calculations of this paper 
were carried out by the Geryon-2 supercluster at the Centro de Astro-Ingenieria UC. This
work also made extensive use of the DiRAC Data Centric system at Durham University, operated by the 
Institute for Computational Cosmology on behalf of the STFC DiRAC HPC Facility (www.dirac.ac.uk). 
This equipment was funded by BIS National E-infrastructure capital grant ST/K00042X/1, 
STFC capital grant ST/H008519/1, and STFC DiRAC Operations grant ST/K003267/1 and Durham University. 
DiRAC is part of the National E-Infrastructure.

\bibliographystyle{mnras}
\bibliography{ref}

\bsp	
\label{lastpage}

\end{document}